\documentclass[useAMS,usenatbib]{mn2e}

\usepackage{natbib}
\usepackage{graphicx}

\newcommand{\beq}{\begin{equation}}
\newcommand{\eeq}{\end{equation}}
\newcommand{\beqa}{\begin{eqnarray}}
\newcommand{\eeqa}{\end{eqnarray}}

\usepackage{epstopdf}
\title[Magnetohydrodynamic-Gravity Waves and Vortices in the Solar Atmosphere]
  {Three-dimensional numerical simulation of magnetohydrodynamic-gravity waves and vortices in the solar atmosphere}
\author[Murawski et al.]
  {K. Murawski$^1$,  I. Ballai$^{2}$, A.K. Srivastava$^{3}$ and D. Lee$^{4}$ \\
  $^1$
      Group of Astrophysics, 
      University of Maria Curie-Sk{\l}odowska, 
      ul. Radziszewskiego 10, 
      20-031 Lublin, Poland \\
      $^2$Solar Physics and Space Plasma Research Centre, Department of Applied Mathematics, The University of Sheffield, Sheffield, UK, S3 7RH \\
  $^3$Aryabhatta Research Institute of Observational Sciences (ARIES), Manora Peak, Nainital-263 129, Uttarakhand, India \\
  $^4$ASC/Flash Center, The University of Chicago, 5640 S. Ellis Ave, Chicago, IL 60637, USA }
\date{Released 2013 Xxxxx XX}

\pagerange{\pageref{firstpage}--\pageref{lastpage}} \pubyear{2013}

\def\LaTeX{L\kern-.36em\raise.3ex\hbox{a}\kern-.15em
    T\kern-.1667em\lower.7ex\hbox{E}\kern-.125emX}

\begin{document}

\label{firstpage}

\maketitle

\begin{abstract}
With the adaptation of 
the FLASH
code 
we simulate 
mag\-ne\-to\-hydro\-dy\-na\-mic--gravity waves and 
vortices 
as well as their response in the magnetized three--dimensional 
(3D) 
solar atmosphere at different heights to understand 
the localized energy transport processes. 
In the solar atmosphere strongly structured 
by gravitational and magnetic 
forces, we launch a localized velocity pulse (in horizontal and vertical components) 
within a bottom layer of 3D solar atmosphere modeled by initial VAL-IIIC conditions, which 
triggers waves and vortices. 
The rotation direction of vortices depends on the orientation of an initial perturbation. 
The vertical driver generates magnetoacoustic-gravity waves which result in oscillations of the transition region, and  
it leads to the eddies with their symmetry axis oriented vertically. 
The horizontal pulse excites all magnetohydrodynamic-gravity waves 
and horizontally oriented eddies. 
These waves propagate upwards, penetrate the transition region, and enter the solar corona. 
In the high-beta plasma regions the magnetic field lines move with the plasma and the temporal evolution show that they swirl with eddies. We estimate the energy fluxes carried out by the waves in the magnetized solar atmosphere and conclude that
such wave dynamics and vortices may be significant in transporting the energy to sufficiently balance the 
energy losses in the localized corona. Moreover, the structure of the transition region highly affects such energy transports,
and causes the channeling of the propagating waves into the inner corona.
\end{abstract}

\begin{keywords}
 Sun: atmosphere -- (magnetohydrodynamics) MHD -- waves.
\end{keywords}

\section{Introduction}
One of the fundamental goals of solar physics research is to explore and understand
dynamical phenomena (waves and transients) in the magnetized plasma of the 
Sun's atmosphere. Space--born observations, and advanced 
multi--di\-men\-sio\-nal numerical models seem to have captured the essentials of coupled dynamics 
of solar plasma and its embedded magnetic field that collectively generates the 
variety of waves in the magnetohydrodynamic (MHD) threshold 
({\it e.g.} Alfv\'en, slow and fast magnetoacoustic) as well as transients 
(vortices, jets, spicules, etc) at diverse spatio-temporal scales. 
The complete understanding of these processes requires to increase the observational and
theoretical efforts, and in particular the intellectual inquiry into fundamental processes of
strongly magnetized and gravitationally stratified plasmas.  
Such MHD waves, instabilities, and dynamical processes in solar and other astrophysical plasma received special attention in the last decades since such effects 
may play an important role in triggering large-scale eruptions, heating, and formation of super--sonic winds. The observed MHD waves and 
oscillations in solar plasmas drove an unprecedented advancement in the context of remote diagnostics of densities, 
magnetic field strength and structure, temperatures, etc. within the framework of MHD seismology 
({\it e.g.} for a review, see Nakariakov, Verwichte, 2005, and references therein). 

High resolution observations of dynamical phenomena occurring on various spatial and temporal scales of the last decade enabled the development of 
complex and realistic numerical/analytical models of waves and oscillations in solar atmospheric plasmas, 
refining MHD seismological methods employed to diagnose remotely the thermal state of the plasma, 
magnitude and sub-resolution structure of the magnetic field, etc. ({\it e.g.} O'Shea {\it et~al.}, 2007; Selwa {\it et~al.}, 2007; Gruszecki {\it et~al.}, 2008; 
Srivastava {\it et~al.}, 2008; Pascoe {\it et~al.}, 2009; 
Andries {\it et~al.}, 2009; Srivastava and Dwivedi, 2010a; Kumar {\it et~al.}, 2011, and references cited therein). 

Recently, at comparatively smaller spatio-temporal scales, the different kind of photospheric/chromospheric 
vortex motions are observed both from space- and ground-based instruments, 
by which the associated magnetic flux-tubes also transfer the responses to upper
transition region and corona in form of waves ({\it e.g.} torsional modes) and plasma motions 
at various spatio-temporal scales ({\it e.g.} 
Jess {\it et~al.}, 2009, and references cited therein). 
Such photospheric vortex motions are also considered as an exciter 
of the magnetohydrodynamic waves ({\it e.g.} Murawski {\it et~al.}, 2013a), while they are inherently originated at the photosphere 
in the granular/intergranular regions (Shelyag {\it et~al.}, 2011) most likely due to the convectively driven flows (Bonet {\it et~al.}, 2008). 
However, the detailed information of the generation of photospheric vortices and their connections with the upwardly launched MHD waves 
as well as plasma jets are still not fully known, and it became the topic of great interest of the researchers recently 
({\it e.g.}  Su {\it et~al.}, 2012, and references cited therein).

The modeling aspects of the vortex motions and their responses in the upper 
solar atmosphere in form of wave and plasma dynamics is very challenging, though the topic of great interest in recent days ({\it e.g.} Murawski {\it et~al.}, 2013a).
Due to the variation of physical conditions throughout the propagation and the change in the dominant 
buoyancy, modeling the evolution of such processes through the entire solar atmosphere is 
associated often with the simplified assumptions by reducing the complexity and focusing on few physical aspects. Numerical investigations are probably 
the only way to use as many assumptions as possible to determine the true nature of the evolution of such motions and their upper atmospheric responses. Despite considerable advance 
in this field of research, there are still several complications connected to how well a numerical code can cope with the realistic profiles of physical parameters.  
The MHD theory is the simplest conceivable model describing the macroscopic behavior and dynamics of plasmas and despite some applicability limitations, it offers a reasonable framework for studying dynamics in the solar atmosphere. At the same time, the MHD equations are strongly nonlinear and therefore their solutions usually require approximate analytical methods ({\it e.g.} multiple scaling or reductive perturbative methods) 
or numerical treatment. Godunov-type methods ({\it e.g.} Murawski and Lee, 2012) are one of several numerical techniques available to solve the MHD equations, 
({\it e.g.} Murawski and Lee, 2011). These methods are simple to implement, easily adaptable to complex geometries, and well suited to handle nonlinear terms. 
Classical numerical schemes of at least second-order accuracy generate polluting oscillations which may result in non--monotonic solutions 
({\it e.g.} Murawski, 2011). Lower-order schemes such as the upwind method (Godunov, 1959)
are usually monotonic, but they are too dissipative, leading to unacceptable solutions 
(Murawski and Lee, 2012). 
In the past, most of the numerical schemes have adopted artificial viscosity to reduce the numerically induced oscillations at 
steep spatial profiles (Stone and Norman, 1992). However, recent experiments with high-order numerical codes proved 
that they are superior in resolving complex profiles which are present in the solar atmosphere (Murawski and Lee, 2012). 

The aim of this paper is to discuss impulsively generated magnetohydrodynamic-gravity waves in low atmospheric layers and their propagation in the magnetised solar atmosphere. 
We present a three--dimensional (3D) numerical model of the solar atmosphere (see Sect.~2) 
and, as a particular application of this model, we study the morphology of the vortices resulted from localized initial velocity pulses. 
One of the major objectives of this paper is also to understand the energy transport by the evolved MHD waves in the different layers 
of the solar atmosphere to balance the localized energy losses. 
This paper is completed by summary of the main results. 
\section{Numerical model of the solar atmosphere}
We consider a gravitationally--stratified solar atmosphere that is described by the set of
ideal, adiabatic, 3D MHD equations given as
\beqa
\label{eq:MHD_rho}
{{\partial \varrho}\over {\partial t}}+\nabla \cdot (\varrho{\bf V})=0\, ,
\\
\label{eq:MHD_V}
\varrho{{\partial {\bf V}}\over {\partial t}}+ \varrho\left ({\bf V}\cdot \nabla\right ){\bf V} =
-\nabla p+ \frac{1}{\mu}(\nabla\times{\bf B})\times{\bf B} +\varrho{\bf g}
\, ,
\\
\label{eq:MHD_p}
{\partial p\over \partial t} + \nabla\cdot (p{\bf V}) = (1-\gamma)p \nabla \cdot {\bf V}\, ,
\hspace{3mm}
p = \frac{k_{\rm B}}{m} \varrho T\, ,
\\
\label{eq:MHD_B}
{{\partial {\bf B}}\over {\partial t}}= \nabla \times ({\bf V}\times{\bf B})\, ,
\hspace{3mm}
\nabla\cdot{\bf B} = 0\, .
\eeqa
Here ${\varrho}$ denotes 
density, ${\bf V}=[V_{\rm x},V_{\rm y},V_{\rm z}]$ is the flow velocity,
${\bf B}=[B_{\rm x},B_{\rm y},B_{\rm z}]$ is the magnetic field, $p$ gas pressure, $T$ temperature,
$\gamma=1.4$ 
is the adiabatic index, 
${\bf g}=(0,-g,0)$ is gravitational acceleration with its magnitude $g=274$ m s$^{-2}$, 
$m$ is mean particle mass that is specified by mean molecular weight value of $1.24$ 
(Oskar Steiner, private communication), 
and $k_{\rm B}$ is the Boltzmann's constant. 

While writing Eqs.~(\ref{eq:MHD_rho})--(\ref{eq:MHD_B}), we assume the plasma to be a completely ionised and 
collisional single fluid throughout the entire domain and we neglect the non--ideal terms such as viscosity, magnetic diffusivity 
and thermal conduction. The latter may play an important role at the transition region, where temperature gradient 
attains a high value. Additionally, we dropped the non--adiabatic effects such as plasma heating and cooling which are important 
in lower regions of the solar atmosphere. However, our objective is to build up our models 
incrementally, with a clear focus on the underlying physical processes at each step. 
Moreover, all neglected effects are not expected to alter the general behaviour 
of waves and vortex plasma motions which will experience amplitude attenuation in non--ideal and non--adiabatic plasma. 
From this point of view our model, described by Eqs.~(\ref{eq:MHD_rho})--(\ref{eq:MHD_B}), seems to be justified. 
Further simplification will be assumed in the configuration of the equilibrium as we will consider a simple vertical magnetic field embedded in realistic solar atmosphere.
\subsection {The equilibrium configuration}
\begin{figure}
\begin{center}
\includegraphics[scale=0.45, angle=0]{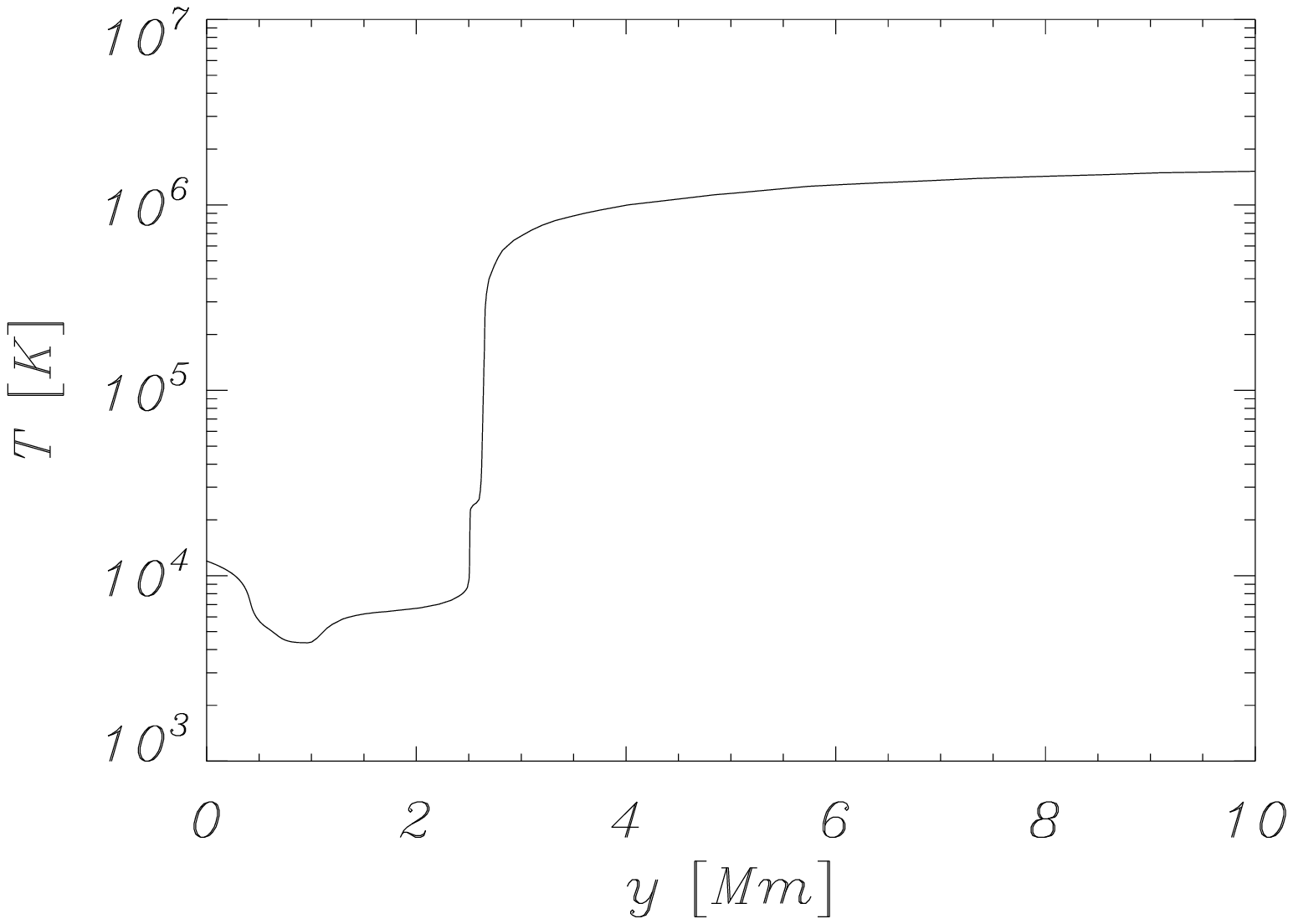}
\includegraphics[scale=0.45, angle=0]{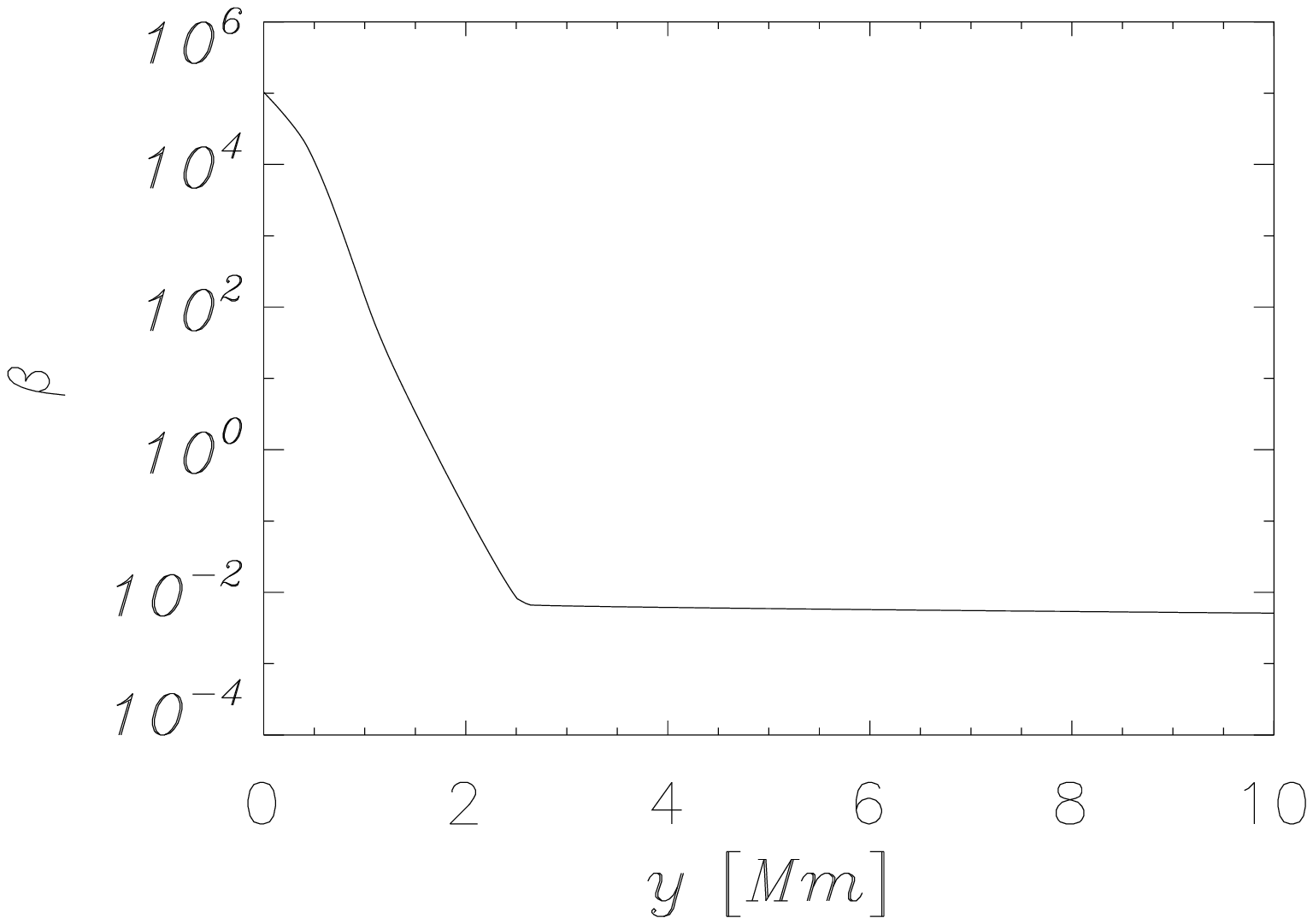}
\caption{\small
Equilibrium profiles of temperature (top panel) and plasma $\beta$ (bottom panel). 
}
\label{fig:initial_profile}
\end{center}
\end{figure}

We assume that the solar atmosphere is in static equilibrium (${\bf V}_{\rm e}={\bf 0}$). 
We express the equilibrium magnetic field (${\bf B}_{\rm e}$) as a constant and vertical magnetic field, i.e.
\beq
{\bf B}_{\rm e} = B_{\rm 0}{\bf \hat y}\ , 
\eeq
with a constant strength 
of $\approx 23$ Gauss that is close a typical magnetic field strength in the quiet Sun (V\"ogler and Sch\"ussler, 2007). 
Here the subscript '${\rm e}$' denotes equilibrium quantities. 

As a result, the equilibrium of our model is not affected by magnetic forces and the equilibrium is reached via the balance of the pressure gradient and gravitational force, i.e. 
\begin{equation}
\label{eq:B}
\varrho_{\rm e} {\bf g} -\nabla p_{\rm e} = {\bf 0}\, .
\end{equation}
Note that all thermodynamical equilibrium quantities have only a vertical coordinate dependence. 

Having specified the equilibrium magnetic field, we can use the momentum equation to find the equilibrium gas pressure and mass density profiles as 
\beqa
\label{eq:pres}
p_{\rm e}(y)=p_{\rm 0}~{\rm exp}\left[ -\int_{y_{\rm r}}^{y}\frac{dy^{'}}{\Lambda (y^{'})} \right]\, ,\hspace{3mm}
\label{eq:eq_rho}
\varrho_{\rm e} (y)=\frac{p_{\rm e}(y)}{g \Lambda(y)}\, ,
\eeqa
where 
\begin{equation}
\Lambda(y) = \frac{k_{\rm B} T_{\rm e}(y)} {mg}
\end{equation}
is the non-constant pressure scale-height, and $p_{\rm 0}$ and $\varrho_{\rm 0}$ 
are the gas pressure and mass density at the reference level that we choose in the solar corona at $y_{\rm r}=10$ Mm. 
Note that $y=0$ corresponds to the base of the photosphere.

We adopt a temperature profile, $T_{\rm e}(y)$, for the solar atmosphere
that is close to the VAL-IIIC atmospheric model of 
\cite{Vernazza_etal}, and smoothly extended into the corona (Fig.~\ref{fig:initial_profile}, top). 
Combining the variation of temperature with height and the hydrostatic pressure balance (Eq.~\ref{eq:pres}) 
we can determine the variation of the equilibrium mass density and gas pressure with height (not shown). 
In this model the temperature reaches $1$ MK at coronal heights and saturates at this level. 
The atmosphere is structured so that the solar photosphere occupies the region $0 < y < 0.5$ Mm, 
the chromosphere is sandwiched between $y=0.5$ Mm and the transition region 
that is located at $y\simeq 2.7$ Mm. 

Once the magnetic field and gas pressure are specified, the plasma beta is given by  
\beq\label{eq:beta}
\beta(y) = \frac{p_{\rm e}}{\frac{B_{\rm e}^2}{2\mu}}\, ,
\eeq 
and it is shown by Fig.~\ref{fig:initial_profile}, bottom panel. 
Note that $\beta$ attains a value of about $10^3$ at $y=0.5$ Mm. 
This value results from magnetic pressure being so small compared to the kinetic pressure. 
For higher values of $y$, $\beta$ falls off to $\approx 0.007$ for $y>2.7$ Mm. This parameter is paramount in the process of investigating dynamics as its value relative to 1 would mean that the dominant forces are of kinetic ($\beta>1$) or magnetic ($\beta<1$) nature.  
\subsection{Perturbations}
The atmospheric magnetic field is continuously perturbed by large-scale dynamical changes that are able to transfer kinetic energy 
({\it e.g.} buffeting due to the granular motion in the photosphere, or flare-driven blast waves in the upper atmosphere, reconnection-driven shocks, etc.). 
We assume that the instigator of changes in the magnetic field has such a nature, however its exact property and nature are not specified. 
Initially (at $t=0$ s) we perturb the above described equilibrium impulsively by a Gaussian pulse either 
in the horizontal, $V_{\rm x}$, or vertical, $V_{\rm y}$, component of velocity, viz.,
\beq\label{eq:perturb}
[V_{\rm x}, V_{\rm y}] (x,y,z,t=0) = 
A_{\rm v} e^{-\frac{x^2+(y-y_{\rm 0})^2+z^2}{w^2}} [s_{\rm x},s_{\rm y}]\, .
\eeq
Here $s_{\rm x}$ and $s_{\rm y}$ attain the discrete values of $0$, $1$. 
The symbol $A_{\rm v}$ denotes the amplitude of the pulse, $y_{\rm 0}$ 
its initial position over the vertical direction, and $w$ is the width of the pulse. We set and hold fixed $A_{\rm v}=3$ km s$^{-1}$, $w=100$ km, and $y_{\rm 0}=500$ km.
\section{Results of numerical simulations}\label{sect:num_res}
\begin{figure}
\begin{center}
\includegraphics[scale=0.5, angle=0]{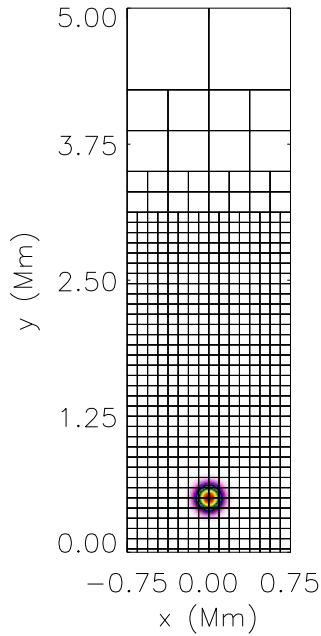}
\caption{\small Numerical blocks with their boundaries (solid lines) and the pulse in velocity of Eq.~(\ref{eq:perturb}) (color maps) 
at $t=0$ s in the vertical plane ($z=0$). 
Maximum velocity (red color) corresponds to 
$3$ km s$^{-1}$.
}
\label{fig:blocks}
\end{center}
\end{figure}
Equations (\ref{eq:MHD_rho})-(\ref{eq:MHD_B}) are solved numerically, using the FLASH code (Lee and Deane, 2009; Lee, 2013). 
This code implements a second-order unsplit Godunov solver (Murawski and Lee, 2011)
with various slope
limiters and Riemann solvers, as well as adaptive mesh refinement. We use the minmod slope limiter and the Roe Riemann solver 
(Murawski and Lee, 2012).
For all considered cases we set the dimension of the simulation box to $(-0.75,0.75)\, {\rm Mm} \times (-0.25,5.75)\times (-0.75,0.75)\, {\rm Mm}$. 
Therefore, the spatial scale of numerical domain is $6$ Mm in the vertical direction and $1.5$ Mm in the horizontal directions. 
At the top and bottom of the numerical domain 
we impose the boundary conditions by fixing in time all plasma quantities
to their equilibrium values, while at the remaining four sides open boundaries are implemented. 
In our simulations we use 
an 
Adaptive Mesh Refinement (AMR) 
grid with a minimum (maximum) level of refinement set to $2$ ($5$). 
The extent of the simulation box in the $y$-direction 
ensures that we catch the essential physics occurring in the solar photosphere-low corona domain 
and minimize the effect of spurious signal reflections from the level $y=5.75$ Mm. 

As each block consists of $8\times 8\times 8$ identical numerical cells, we reach the effective finest spatial resolution of about $30$ km, 
below 
the altitude 
$y=3.25$ Mm. The initial system of blocks is shown in Figure~\ref{fig:blocks}. In the remaining part of the paper we discuss two particular cases of Eq.~(\ref{eq:perturb}): 
(a) vertical perturbation with $s_{\rm x}=0$ and $s_{\rm y}=1$; 
(b) horizontal perturbation, corresponding to $s_{\rm x}=1$, $s_{\rm y}=0$.
\subsection{Vertical perturbation}
\begin{figure}
\begin{center}
\includegraphics[width=7.00cm,height=8.50cm, angle=0]{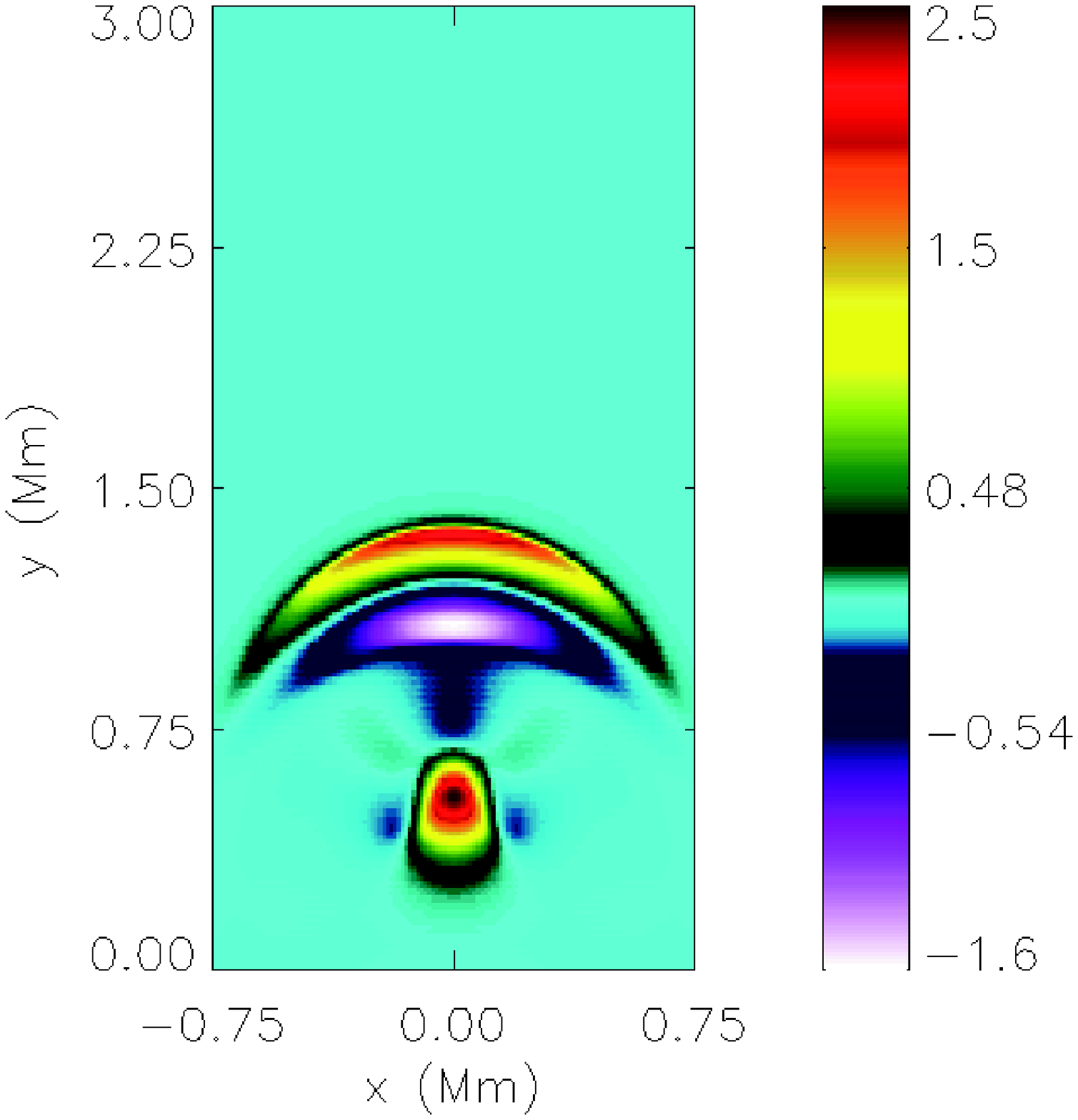}
\includegraphics[width=7.00cm,height=8.50cm, angle=0]{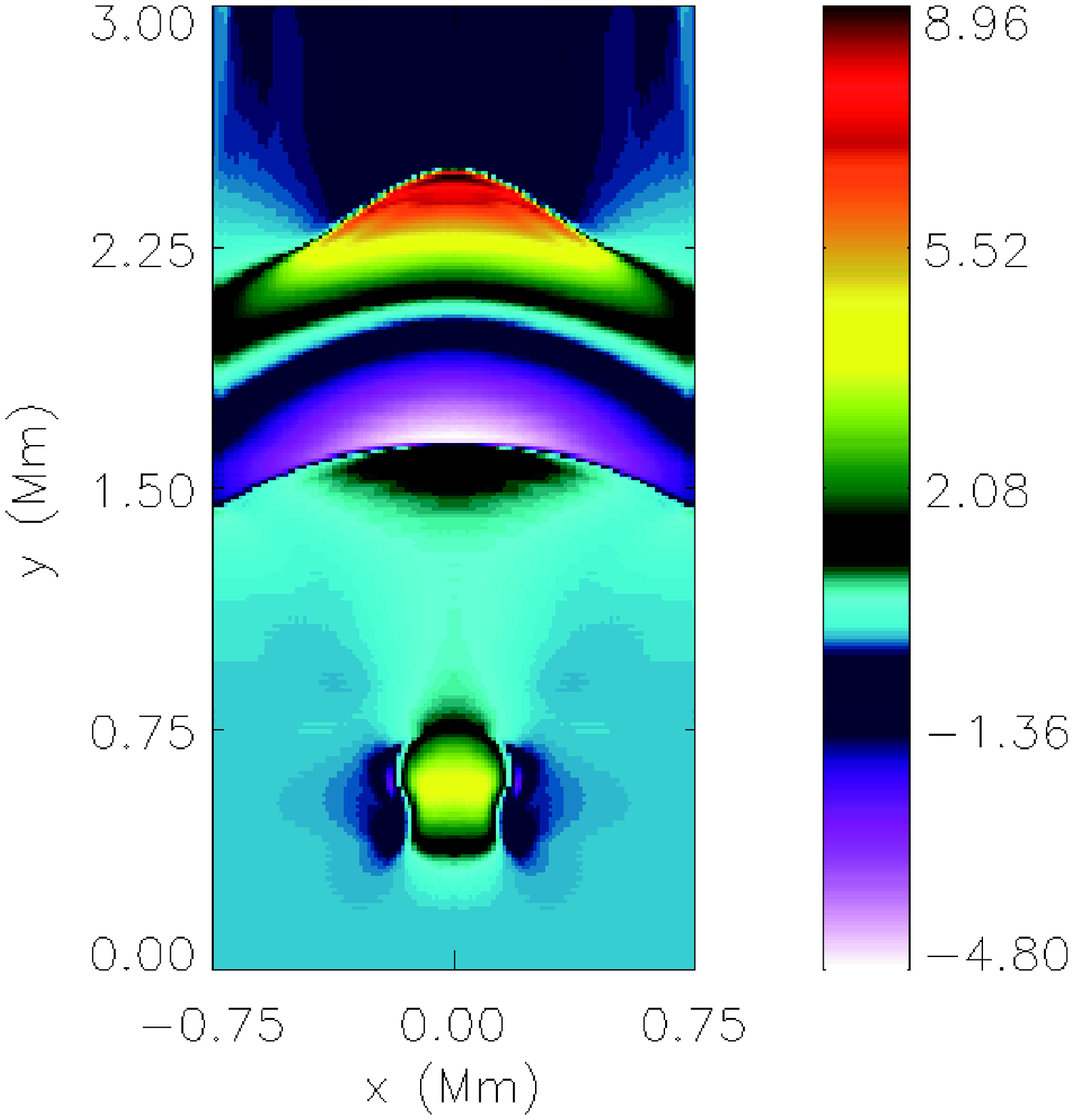}
\caption{\small Temporal snapshots of $V_{\rm y}(x,y,z=0)$ 
at $t=100$ s 
(top) 
and 
$t=200$ s 
(bottom) 
for the case of the vertical perturbation with $s_{\rm x}=0$, $s_{\rm y}=1$ in Eq.~(\ref{eq:perturb}). 
Velocity 
is expressed in units of $1$ km s$^{-1}$. 
The source is situated at $y_0=0.5$ Mm. 
}
\label{fig:vert-Vy}
\end{center}
\end{figure}
In the present section we consider the case of vertical perturbations which are realized by setting $s_{\rm x}=0$ and $s_{\rm y}=1$ 
in Eq.~(\ref{eq:perturb}). Earlier investigations, by e.g. Jain et al. (2011), showed that $p$-modes are able to provide the necessary initial perturbation considered here. 
The initial pulse given by Eq.~(\ref{eq:perturb}) triggers longitudinal magnetoacoustic--gravity waves propagating in the solar atmosphere. 
Figure~\ref{fig:vert-Vy} illustrates vertical component of velocity, $V_{\rm y}$, 
resulting from the initial perturbation. 
The initial pulse splits into counter-propagating waves (Fig.~\ref{fig:vert-Vy}, top panel). 
As the plasma $\beta$ is high at the launching point, the initial pulse excites both fast- and slow-magnetoacoustic-gravity waves 
which are coupled one to each other. 
At $t=100$ s, these waves reached the point $(x=0,y\approx 1.4)$ Mm (Fig.~\ref{fig:vert-Vy}, top), 
spreading quasi-isotropically out of the launching place. 
The red and violet patches move towards the transition region.
The red (violet) patch exhibits the positive (negative) vertical velocity of the perturbations. 
At $t=200$ s the leading wave-front reached the transition region (Fig.~\ref{fig:vert-Vy}, bottom). 

One indication of the nature of these waves is their propagation speed. 
It is found that the average propagation speed is approximately $12$ km s$^{-1}$. 
Due to the increase in temperature and decrease in density, the amplitude of these waves 
grows 
with height 
(see the colour code attached to each figure). 
At $t=200$ s (bottom panel), the fast magnetoacoustic--gravity excited perturbations, initially for high plasma $\beta$ seen 
in the vertical component of velocity, have already penetrated the transition region and entered the solar corona. 

%
\begin{figure}
\begin{center}
\includegraphics[width=7.00cm,height=8.50cm, angle=0]{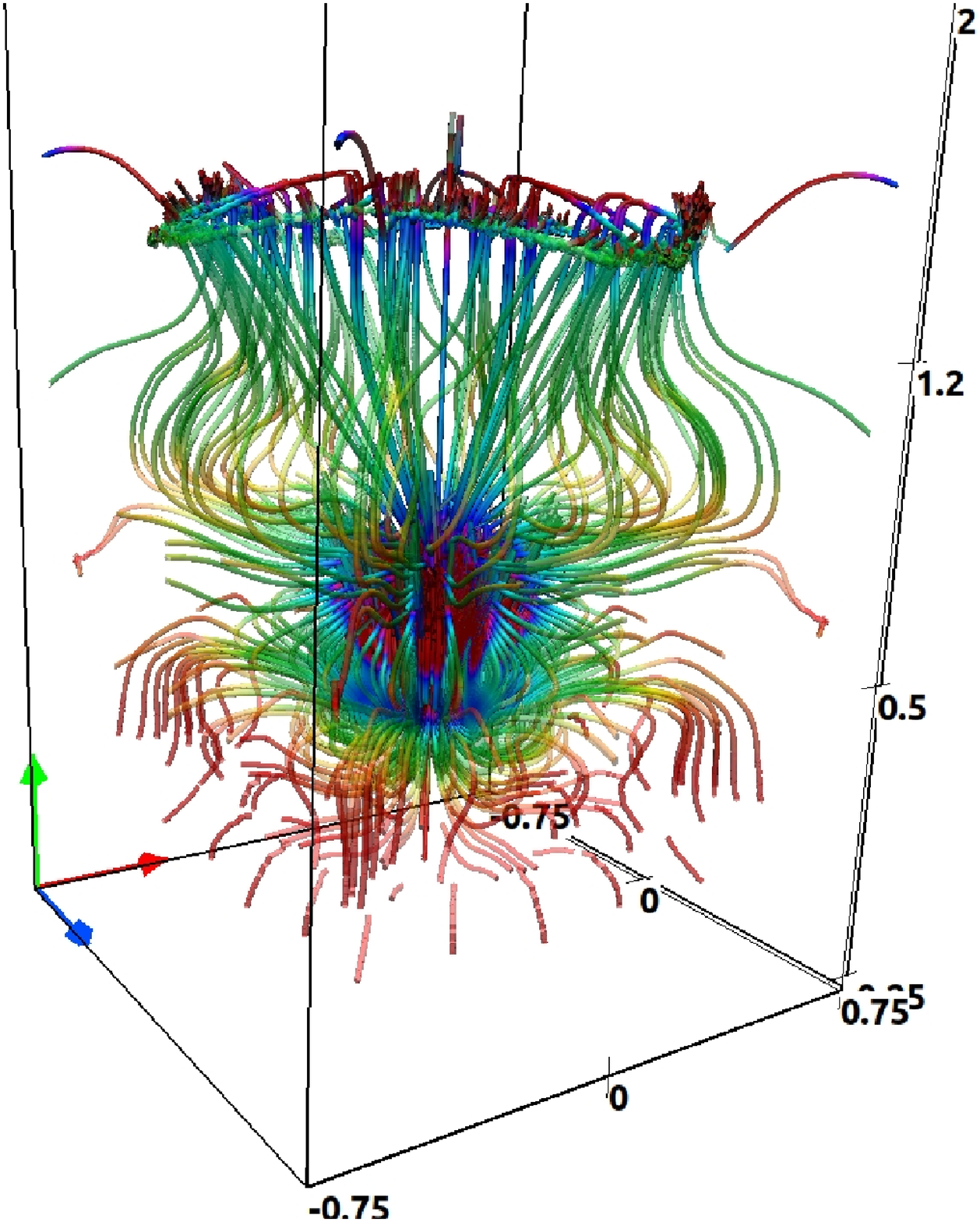}
\includegraphics[width=7.00cm,height=8.50cm, angle=0]{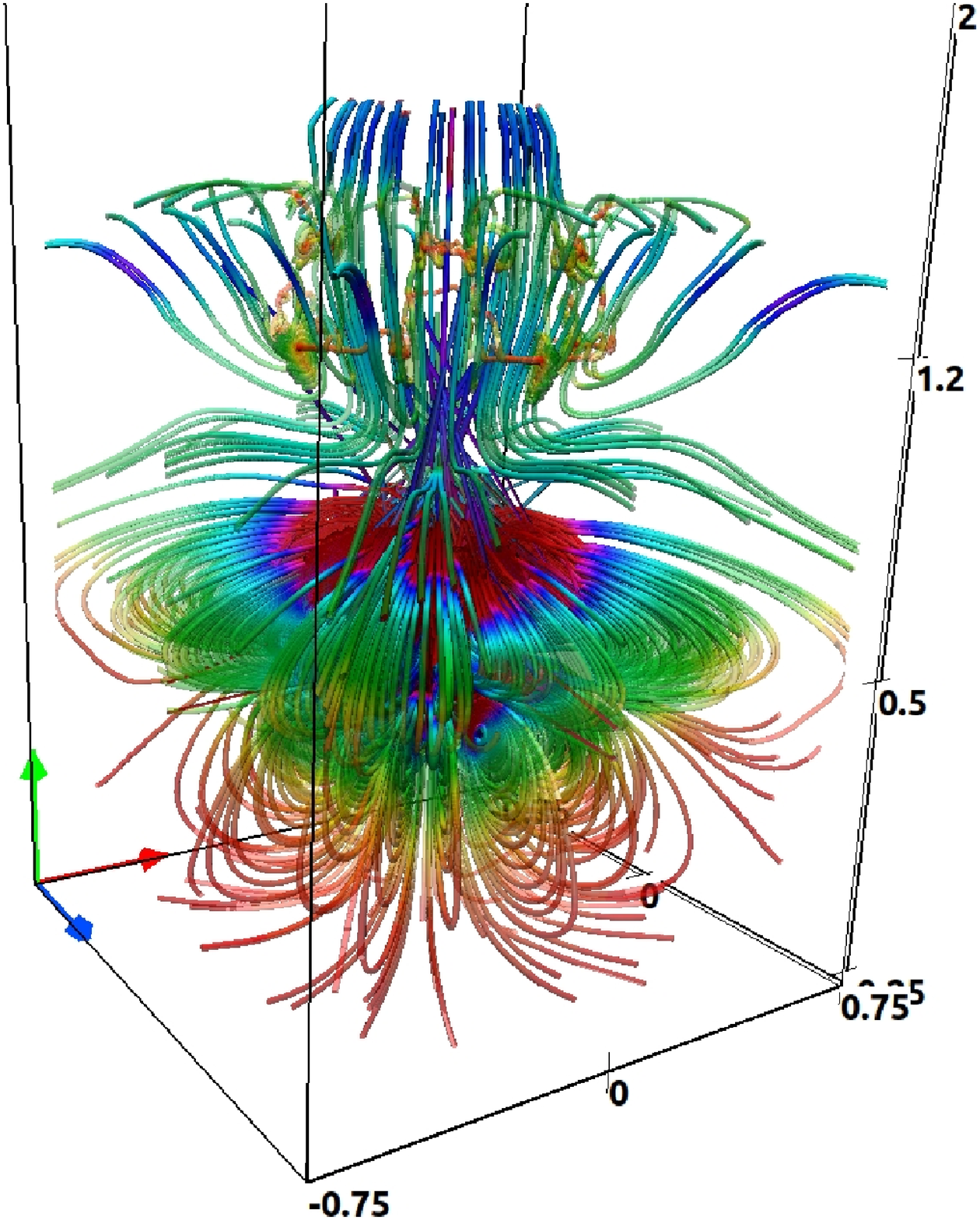} 
\caption{\small 
Temporal snapshots of 
streamlines at 
$t=200$ s 
(top) 
and $t=300$ s 
(bottom) 
for the case of the vertical perturbation with $s_{\rm x}=0$, $s_{\rm y}=1$ in Eq.~(\ref{eq:perturb}). 
Red, green, and blue arrows correspond to the $x$-, $y$-, and $z$-axis, respectively. 
}
\label{fig:vert-V}
\end{center}
\end{figure}
Temporal snapshots of streamlines are displayed at 
$t = 100$ s (Fig.~\ref{fig:vert-V}, top panel) and 
$t = 200$ s (bottom panel). 
These streamlines, defined by the expression 
\beq
\frac{dx}{V_{\rm x}} = \frac{dy}{V_{\rm y}} = \frac{dz}{V_{\rm z}} \, ,
\eeq 
reveal vortices which develop in time into more complex structures with the symmetry axis $y$. 
Such a development of vortices albeit in a 2D uniform magnetic--field case was already reported by 
\cite{Murawski_etal}. 
Vortices were also found to accompany acoustic waves in the case they are triggered by 
velocity pulses launched initially in a uniform, 3D, magnetic-free medium \citep{Mur-Mur-Sch}. 
\begin{figure}
\begin{center}
\includegraphics[width=7.00cm,height=8.50cm, angle=0]{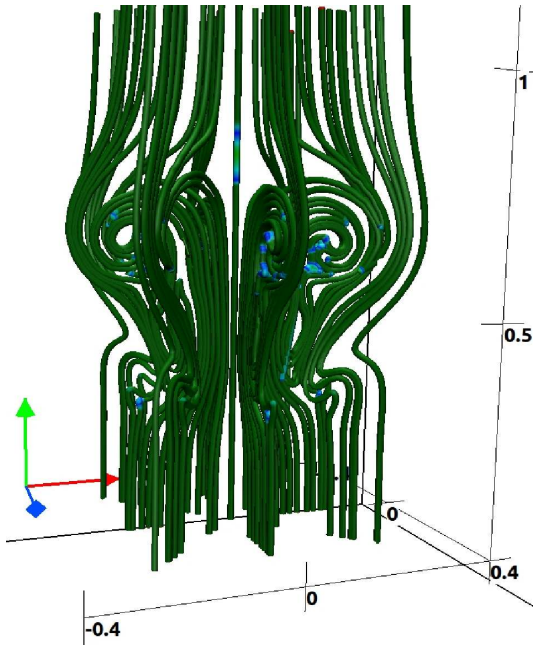}
\caption{\small Temporal snapshots of magnetic-field lines at 
$t=300$ s for the case of the vertical perturbation with $s_{\rm x}=0$, $s_{\rm y}=1$ in Eq.~(\ref{eq:perturb}). 
Red, green, and blue arrows correspond to the $x$-, $y$-, and $z$-axis, respectively. 
}
\label{fig:vert-B}
\end{center}
\end{figure}

It is noteworthy that waves and vortices result in alteration of magnetic field lines 
which are pushed off from the central line, $x=0$ Mm, well seen at $t=300$ s (Fig.~\ref{fig:vert-B}). 
As a result of vortices, which are associated with up-flowing and down-flowing plasma, the central magnetic--field lines 
are looped at $t=300$ s (Fig.~\ref{fig:vert-B}). 
These vortices experience energy cascade into smaller scales, and they are present until the end of our simulation runs. 
The first vortex results from the initial pulse in $V_{\it y}$, which is a characteristic feature of velocity perturbations (Murawski et al. 2013b). 
Such vortices were also theorized by 
{\rm Konkol, Murawski, Zaqarashvili} (2012) in a similar but 2D context. 
The observed vertical vortices 
are located at the height below 1 Mm that possesses a very high plasma beta. Therefore, the diffusion and twisting of the magnetic field with 
the local rotational velocity field of the plasma may yield the simulated vertical vortices in such a weakly magnetized solar atmosphere.
\begin{figure}
\begin{center}
\includegraphics[width=6.00cm,height=8.50cm, angle=0]{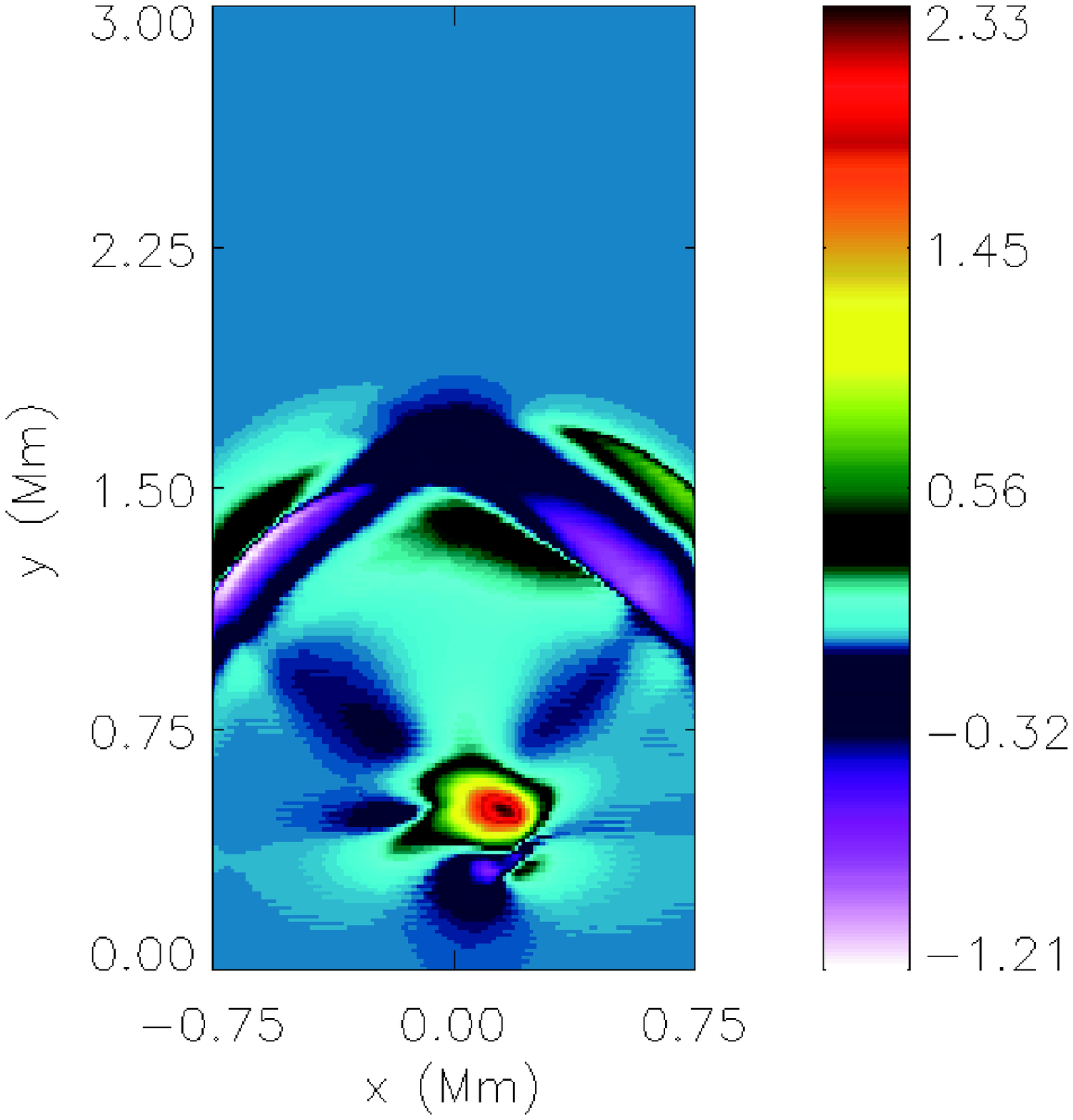}
\includegraphics[width=6.00cm,height=8.50cm, angle=0]{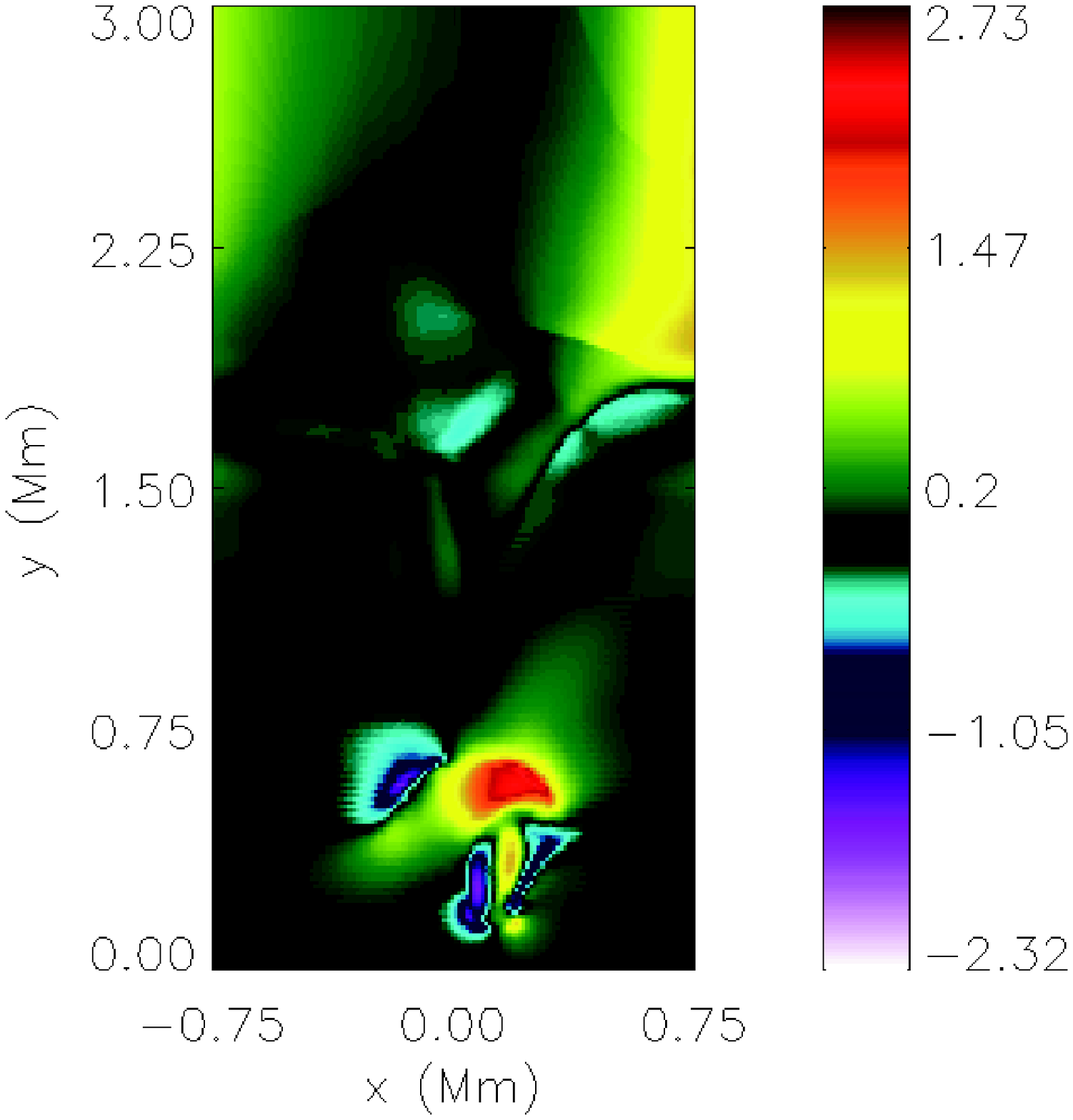}
\caption{\small Temporal snapshots of $V_{\rm x}(x,y,z=0)$ at 
$t=150$ s (top panel) and $t=250$ s (bottom panel) for the case of the horizontal perturbation with $s_{\rm x}=1$, $s_{\rm y}=0$
in Eq.~(\ref{eq:perturb}). 
Velocity is expressed in units of $1$ km s$^{-1}$. 
}
\label{fig:horizon-Vx}
\end{center}
\end{figure}
\begin{figure}
\begin{center}
\includegraphics[width=6.00cm,height=8.50cm, angle=0]{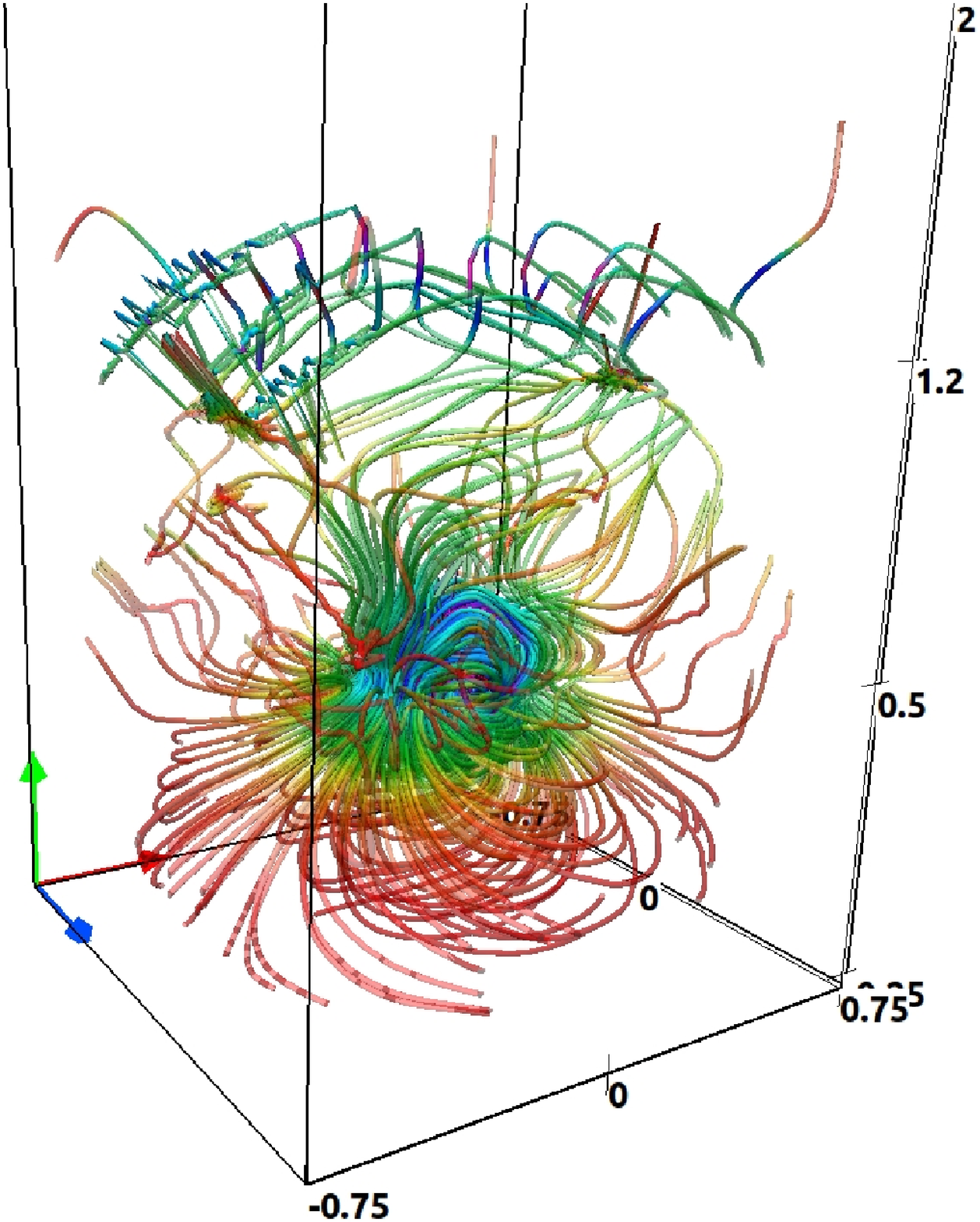}
\includegraphics[width=6.00cm,height=8.50cm, angle=0]{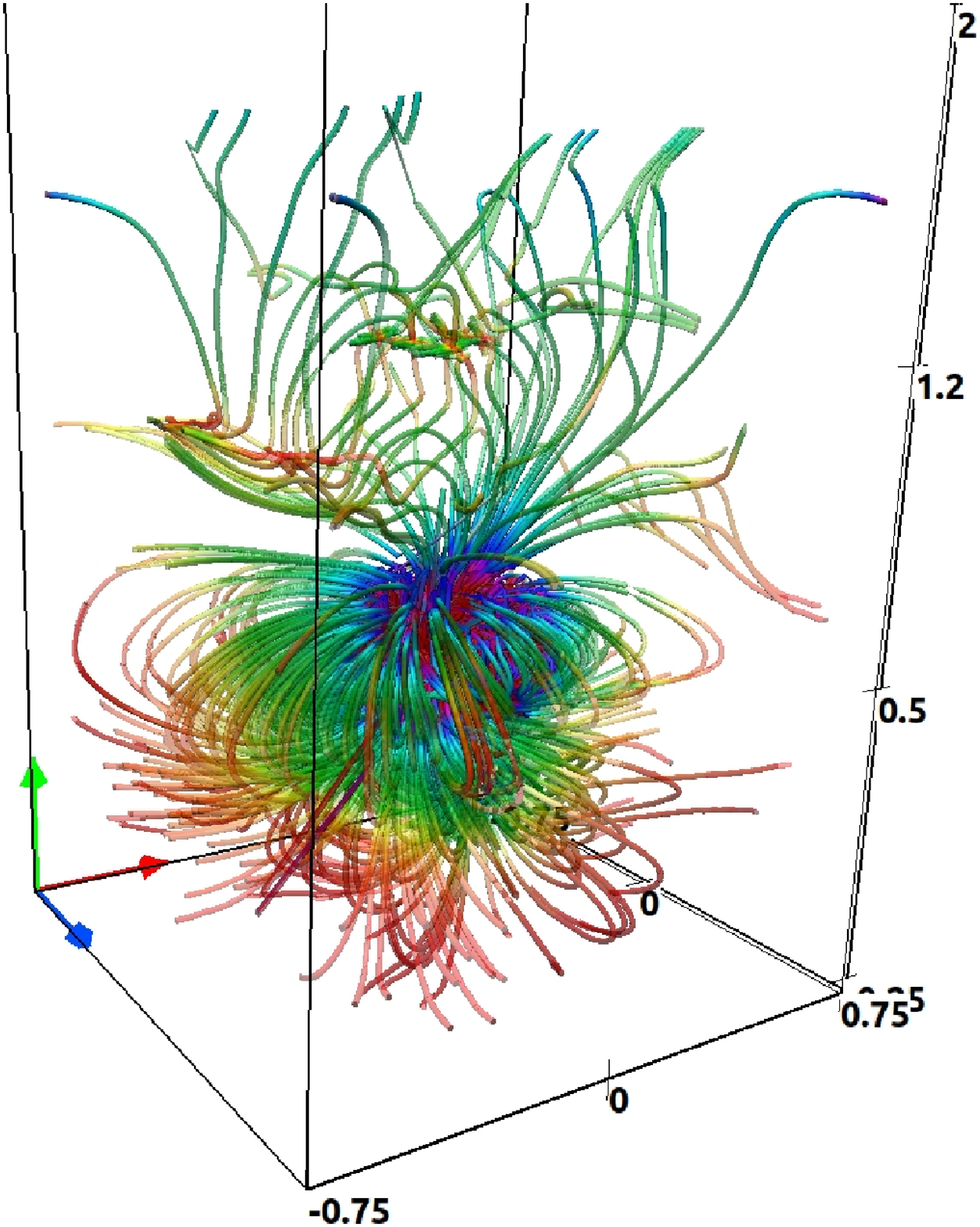}
\caption{\small Temporal snapshots of streamlines at 
$t=150$ s (top panel) 
and $t=250$ s 
(bottom panel) 
for the case of the horizontal perturbation with $s_{\rm x}=1$, $s_{\rm y}=0$
in Eq.~(\ref{eq:perturb}). 
Red, green, and blue arrows correspond to the $x$-, $y$-, and $z$-axis, respectively. 
}
\label{fig:horizV}
\end{center}
\end{figure}
\subsection{Horizontal perturbation}
We discuss here the case of horizontal perturbations which are described by Eq.~(\ref{eq:perturb}) with $s_{\rm x}=1$ and $s_{\rm y}=0$. Drivers of this type are easily found in solar atmosphere as these would correspond to, e.g. the continuous buffeting of magnetic field lines by granular cells and given the nature of the resulting waves, transversal waves of this type are the most easy to trigger. 
Fig.~\ref{fig:horizon-Vx} illustrates spatial profiles of the horizontal velocity components at 
$t=150$ s (top panel) and $t=250$ s (bottom panel). The initial pulse 
excites all magnetohydrodynamic waves which are altered by the gravity field. 
The magnetoacoustic-gravity waves (violet wave-fronts) are seen at $t=150$ s, reaching the level of about $y=1.75$ Mm 
(top panel). The upwardly propagating waves head on towards the transition region 
(bottom panel) and later on they partially reflect from the transition region, 
while their significant part penetrates into the solar corona (bottom panel). 

Similarly to the case of the vertical driver, the horizontal driver generates vortices, 
however, with entirely different inherent physical scenario. 
Temporal snapshots of streamlines are displayed in Fig.~\ref{fig:horizV}. 
At $t = 150$ s (top panel) there are well seen eddies which occupy a region close to the launching place. 
These eddies evolve into more complex rotating structures at later moments of time, 
{\it e.g.} at $t=250$ s (bottom panel). Comparing Figs.~4 and 7 we infer that in the case of the horizontal perturbation 
the resulted eddies exhibit less symmetry and their morphology is more complex than the vortices in the case of 
the initial vertical velocity pulse. 
Note that these eddies are directly triggered by the corresponding initial pulses and they do not result from 
convective instabilities which settle at the launching place at $t\approx 200$ s (not shown). 

\begin{figure}
\begin{center}
\includegraphics[width=6.00cm,height=8.50cm, angle=0]{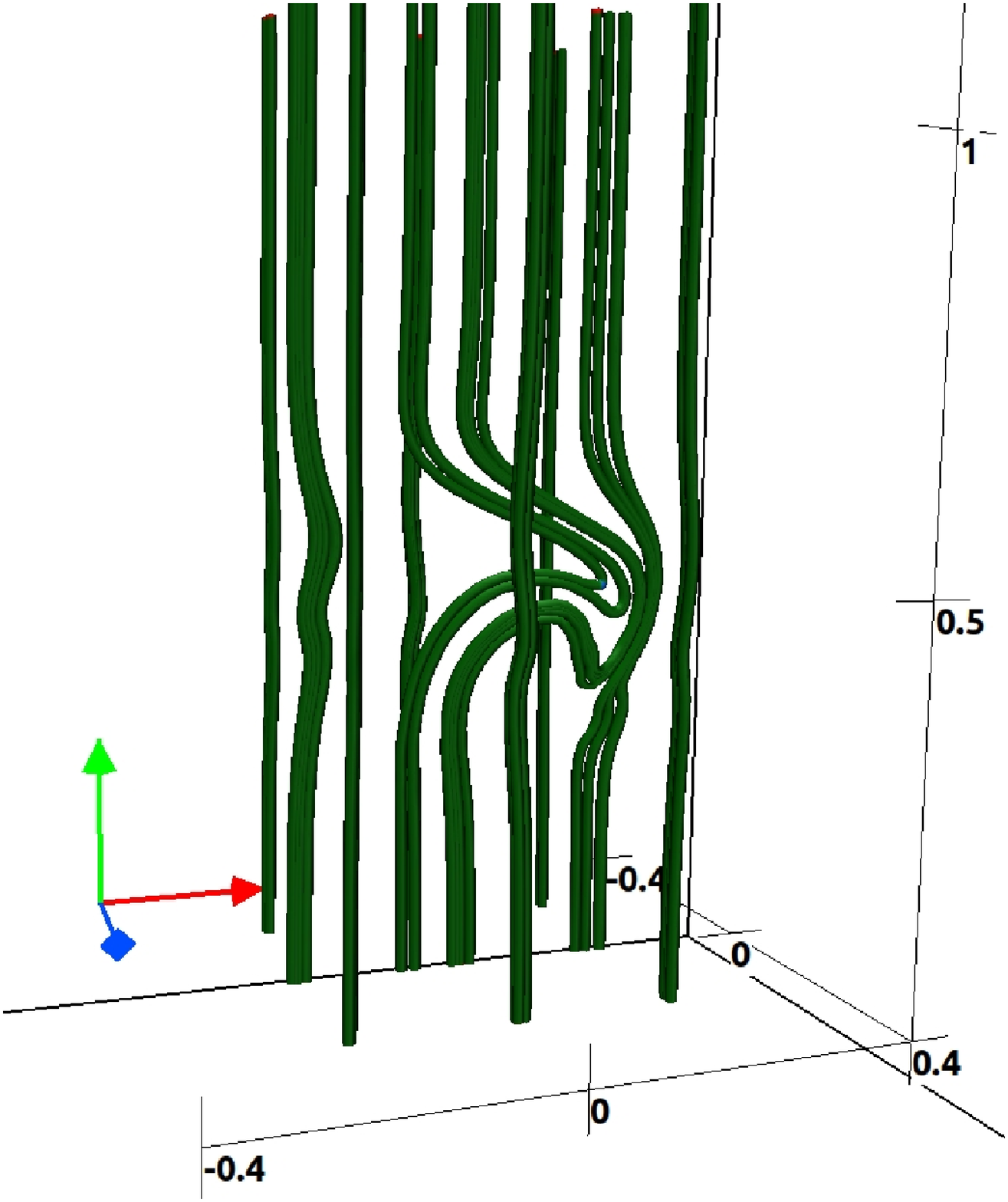}
\includegraphics[width=6.00cm,height=8.50cm, angle=0]{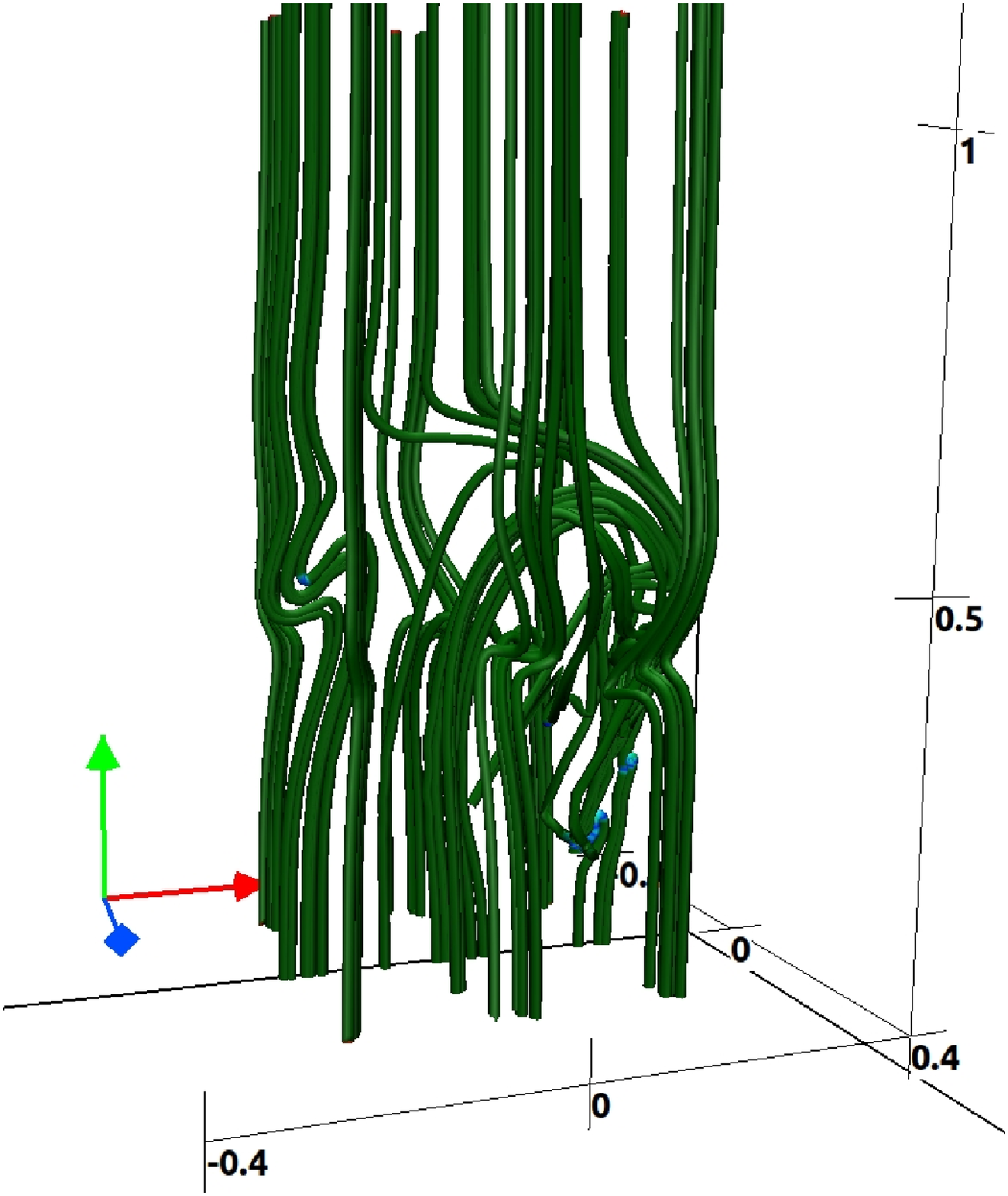}
\caption{\small Temporal snapshots of magnetic-field lines at 
$t=150$ s (top panel) and $t=250$ s (bottom panel) for the case of the horizontal perturbation with $s_{\rm x}=1$, $s_{\rm y}=0$ in Eq.~(\ref{eq:perturb}). 
Red, green, and blue arrows correspond to the $x$-, $y$-, and $z$-axis, respectively. 
}
\label{fig:horiz-B}
\end{center}
\end{figure}
The magnetic field-lines, illustrated in Fig.~\ref{fig:horiz-B}, top panel, are bent concave towards positive values of $x$. 
These curved lines show that the kink wave is excited. However, at later moments of time 
magnetic field lines exhibit more complex braiding that accompany the formation of the vortices (bottom panel). 
The horizontal vortices in strong 
magnetic fields may have morphologically
very different characteristics than the vertical vortices
generated in the weak vertical magnetic fields. In the case of horizontal vortices triggered by convective 
motions or granular buffeting, magnetic tension prevents strong twisting of magnetic field lines, and therefore, none
long-lived rotating structures are appeared. 
The twisting of the field lines that we observe, is located in the high-beta plasma region, therefore, we see the strong 
deformation of the field lines, which follow the velocity field. 
The lifetime of these structures is about $500$ seconds, which is 
comparable 
to the characteristic time required for the flow to propagate through the magnetic structure. 
\subsection{Energy Fluxes Associated with the MHD Wave Propagation and their Comparison}
\begin{figure*}
\begin{center}
\mbox{
\includegraphics[scale=0.3, angle=0]{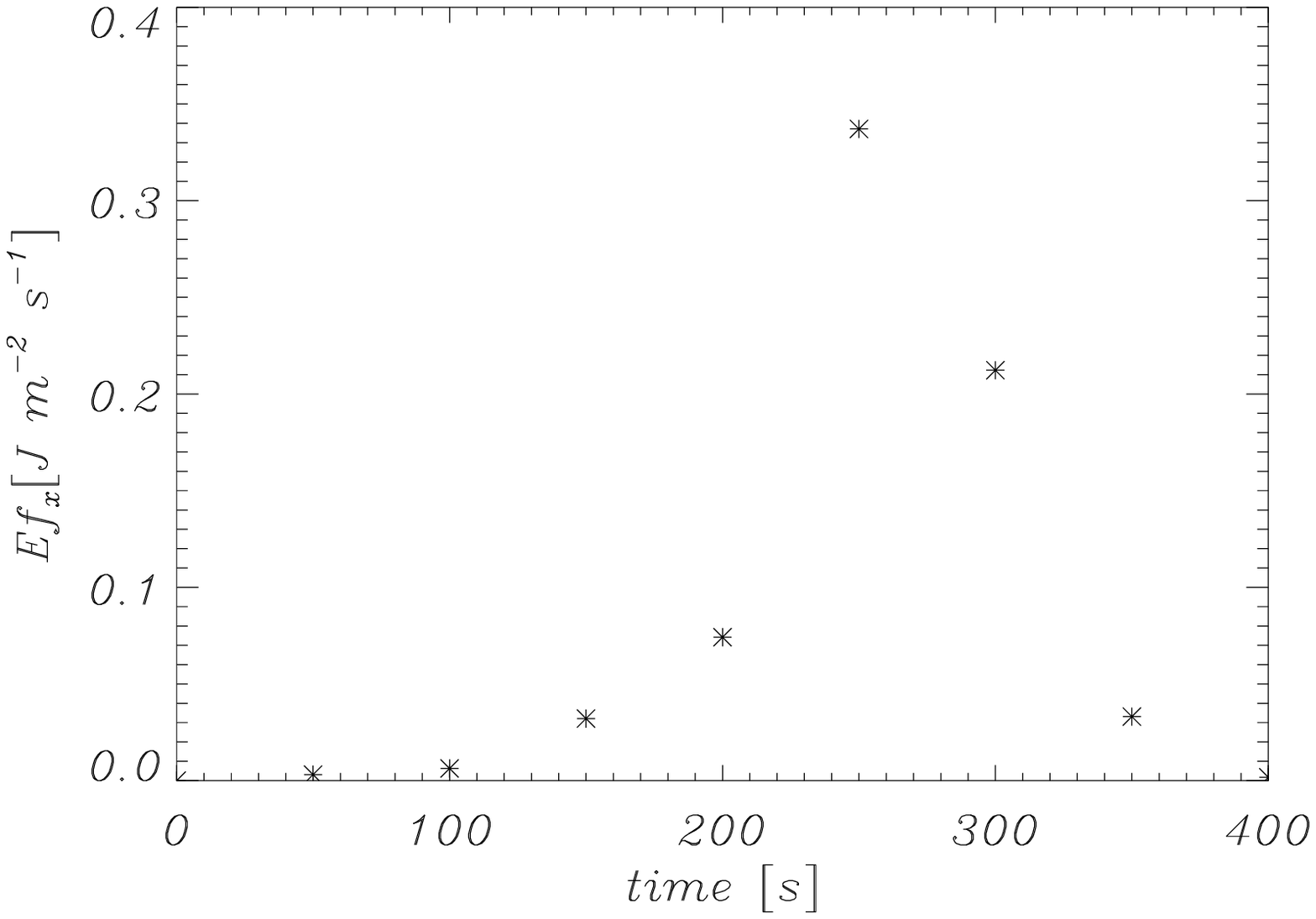}
\includegraphics[scale=0.3, angle=0]{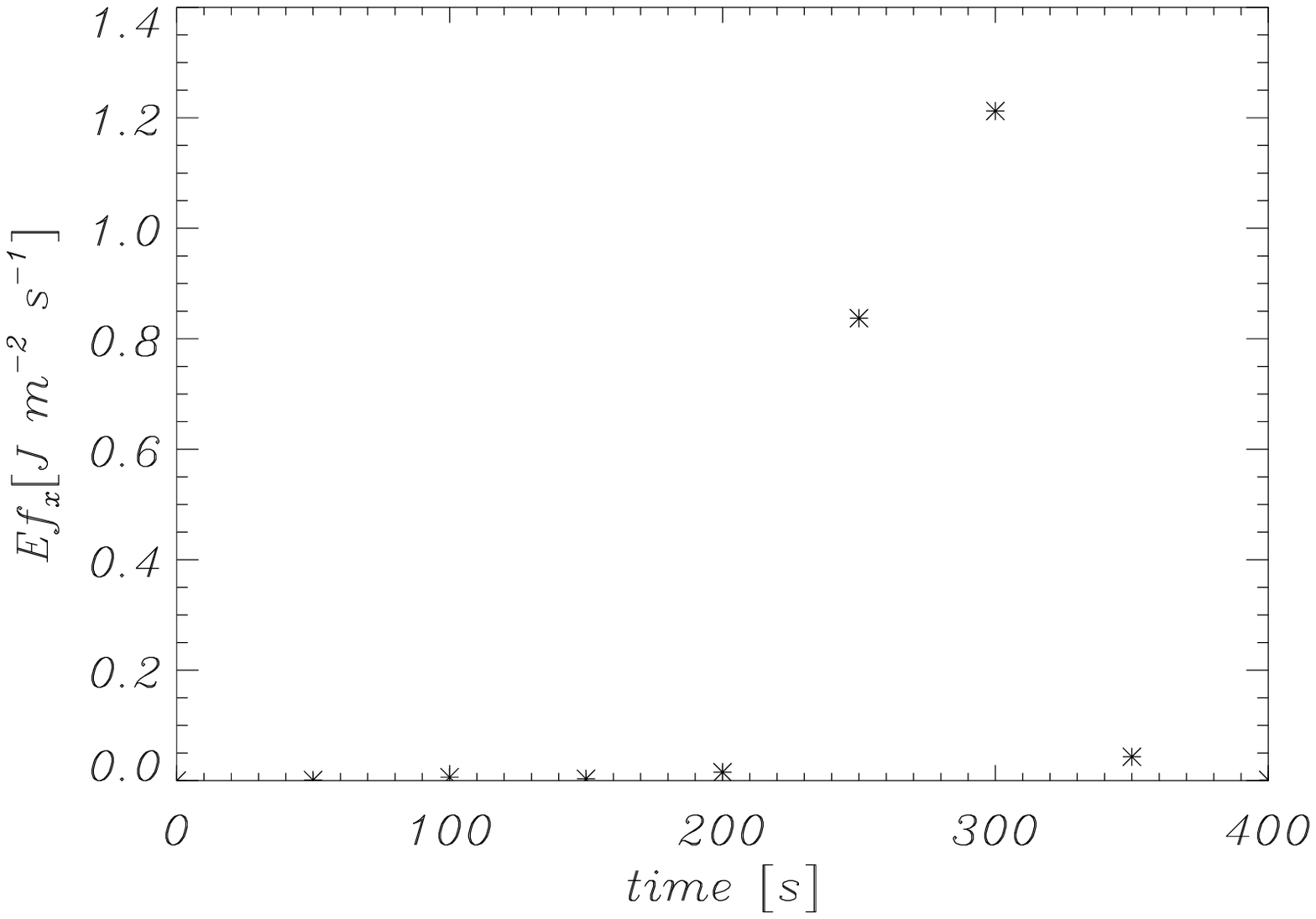}
\includegraphics[scale=0.3, angle=0]{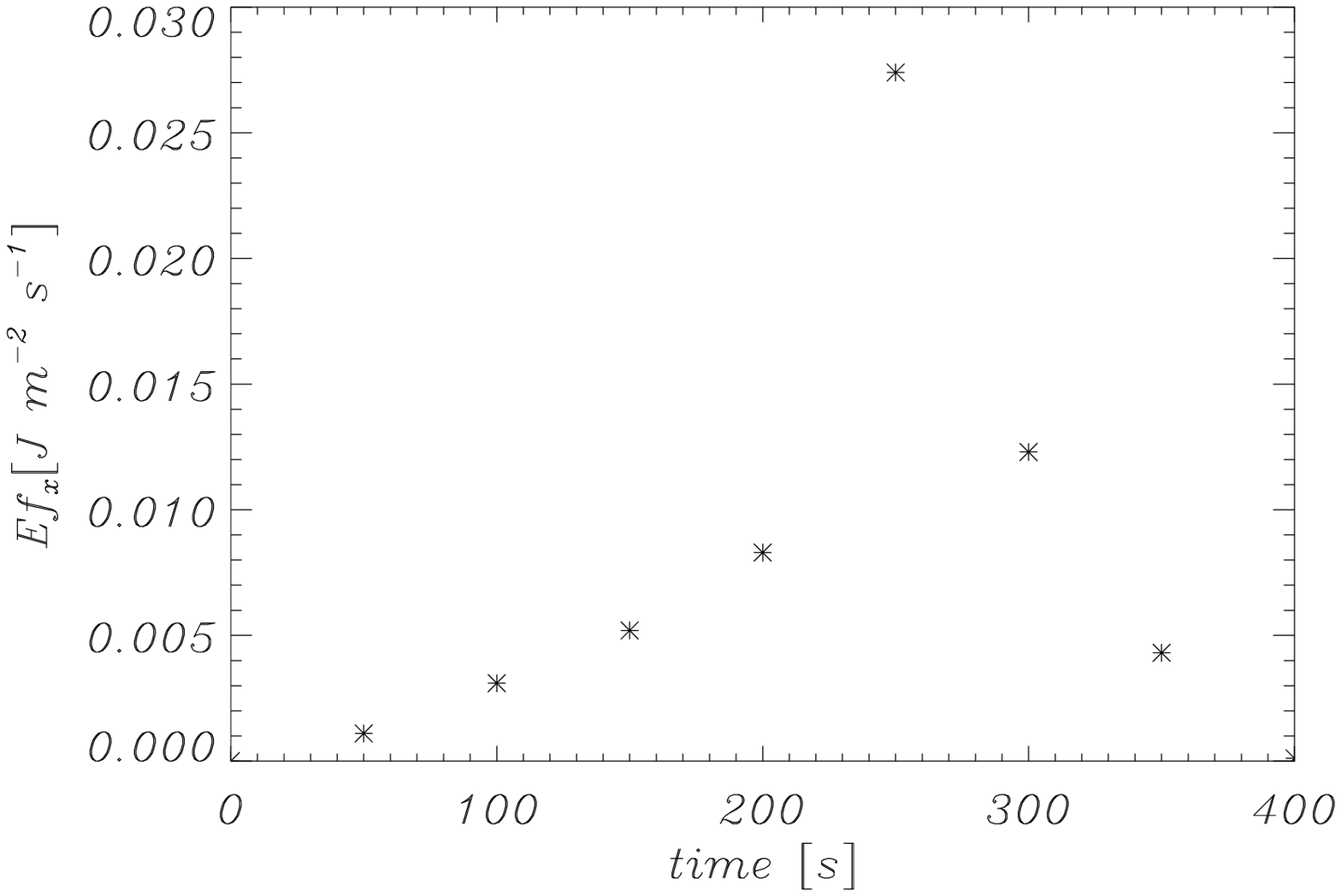}
}
\mbox{
\includegraphics[scale=0.3, angle=0]{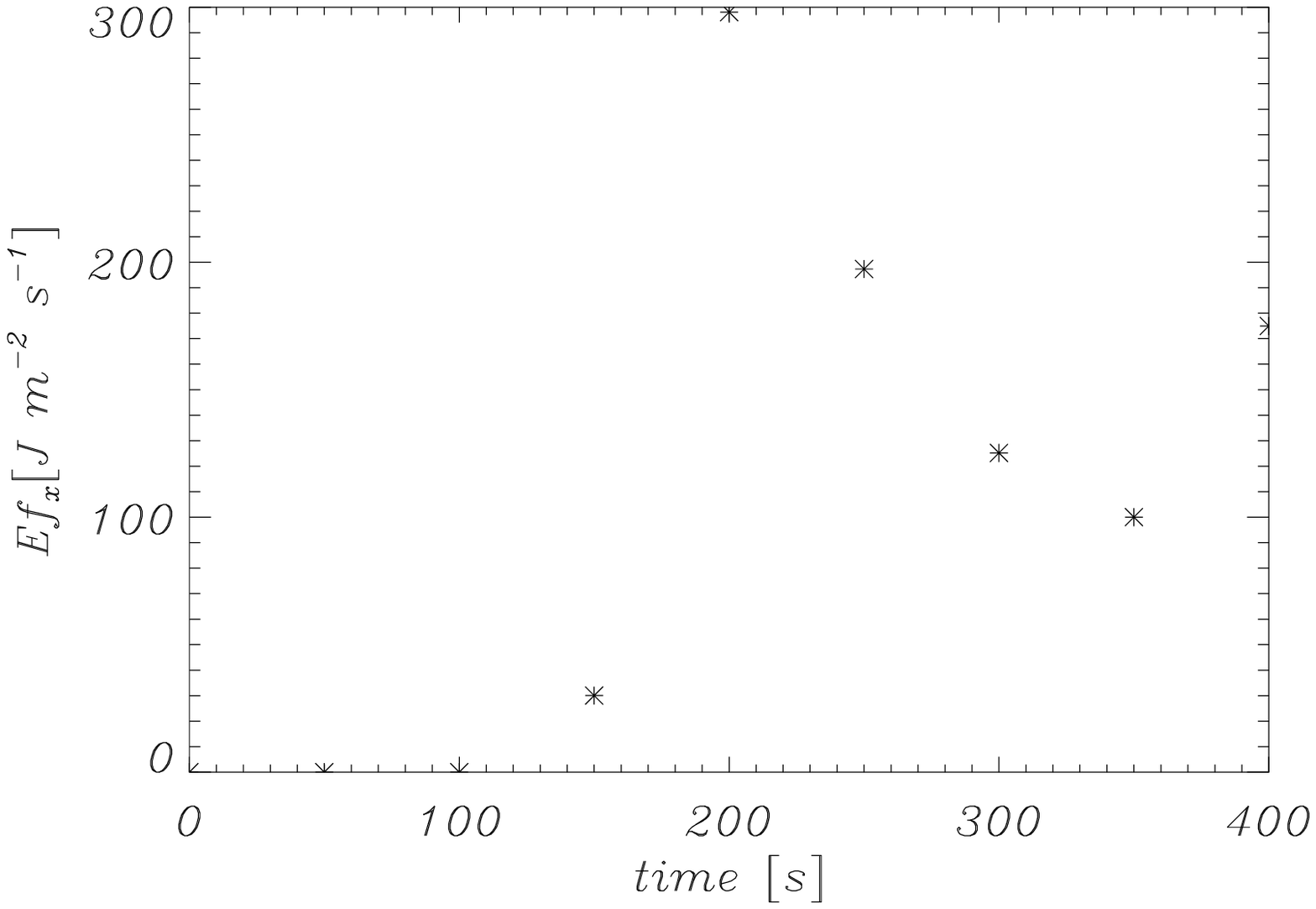}
\includegraphics[scale=0.3, angle=0]{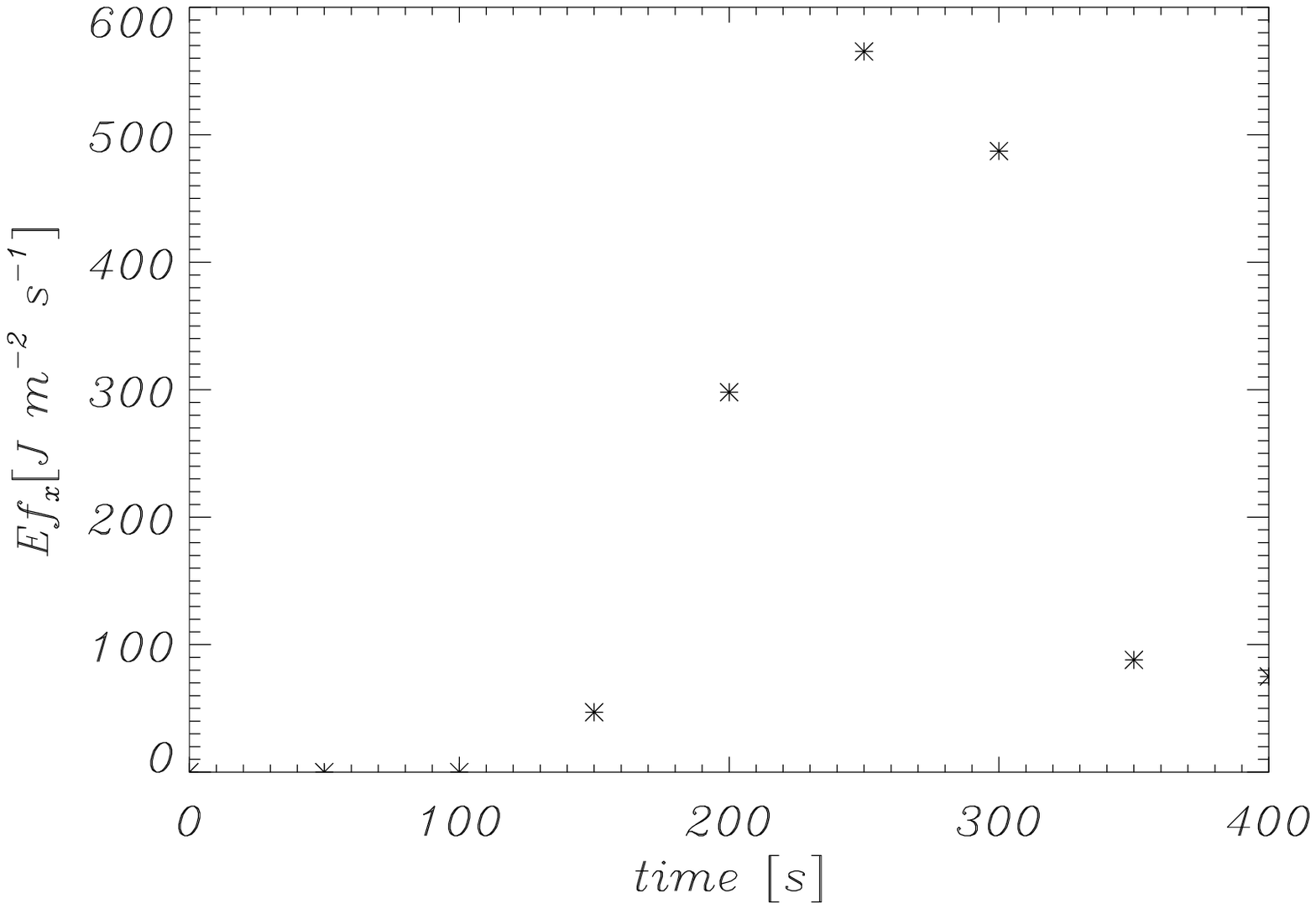}
\includegraphics[scale=0.3, angle=0]{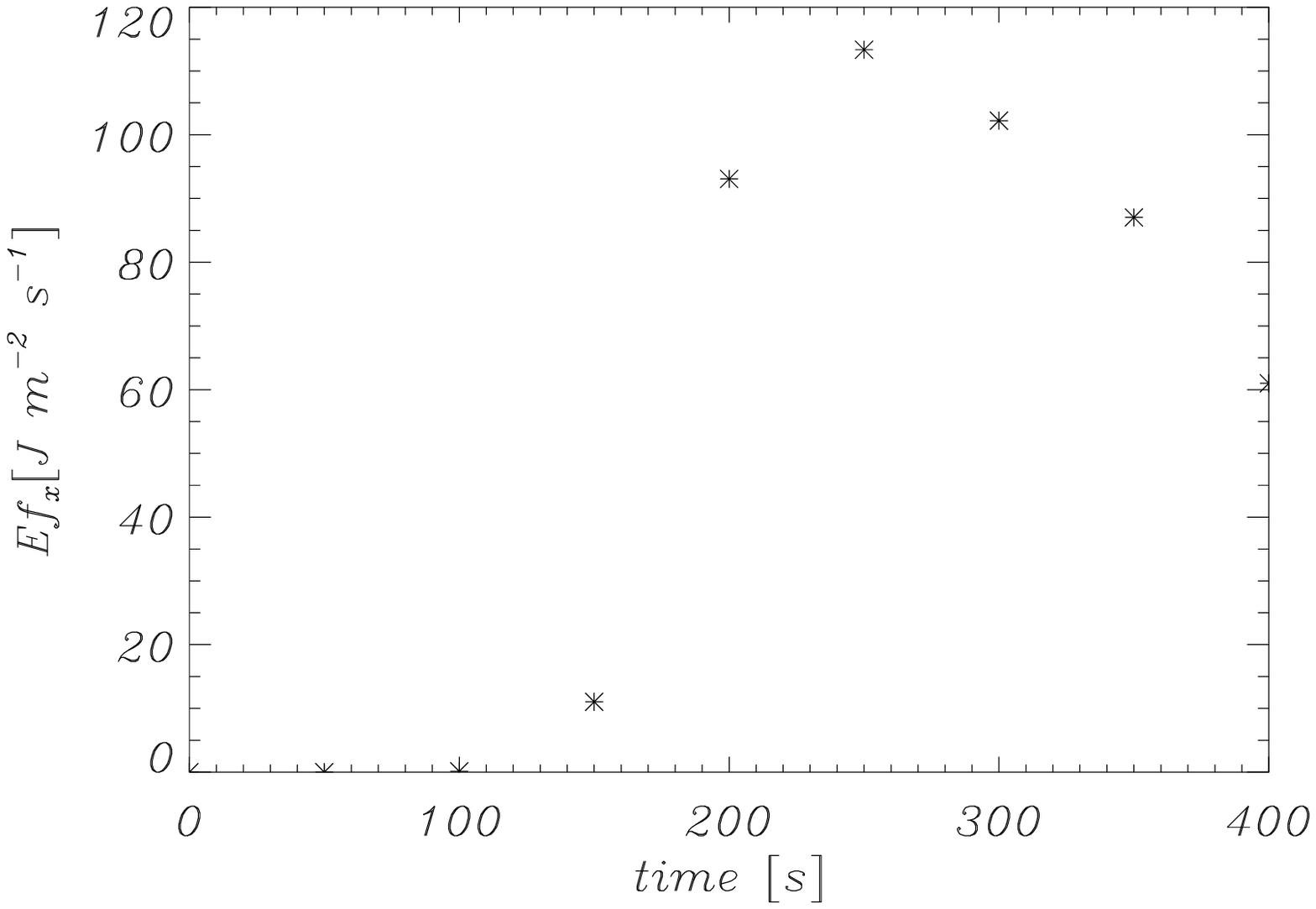}
}
\caption{\small Temporal snapshots of the $x-$ and $y-$components 
of the horizontally averaged energy fluxes, $Ef_{\rm x}$, $Ef_{\rm y}$, evaluated at 
(i)  $y=2.5$ Mm (just below the transition region, left columns), 
(ii) $y=3.0$ Mm (slightly above the transition region, middle columns), and 
(iii) $y=3.5$ Mm (in the inner corona, right columns) 
for the case of the vertical pulse. 
}
\label{fig:vert-Ef}
\end{center}
\end{figure*}

\begin{figure*}
\begin{center}
\mbox{
\includegraphics[scale=0.3, angle=0]{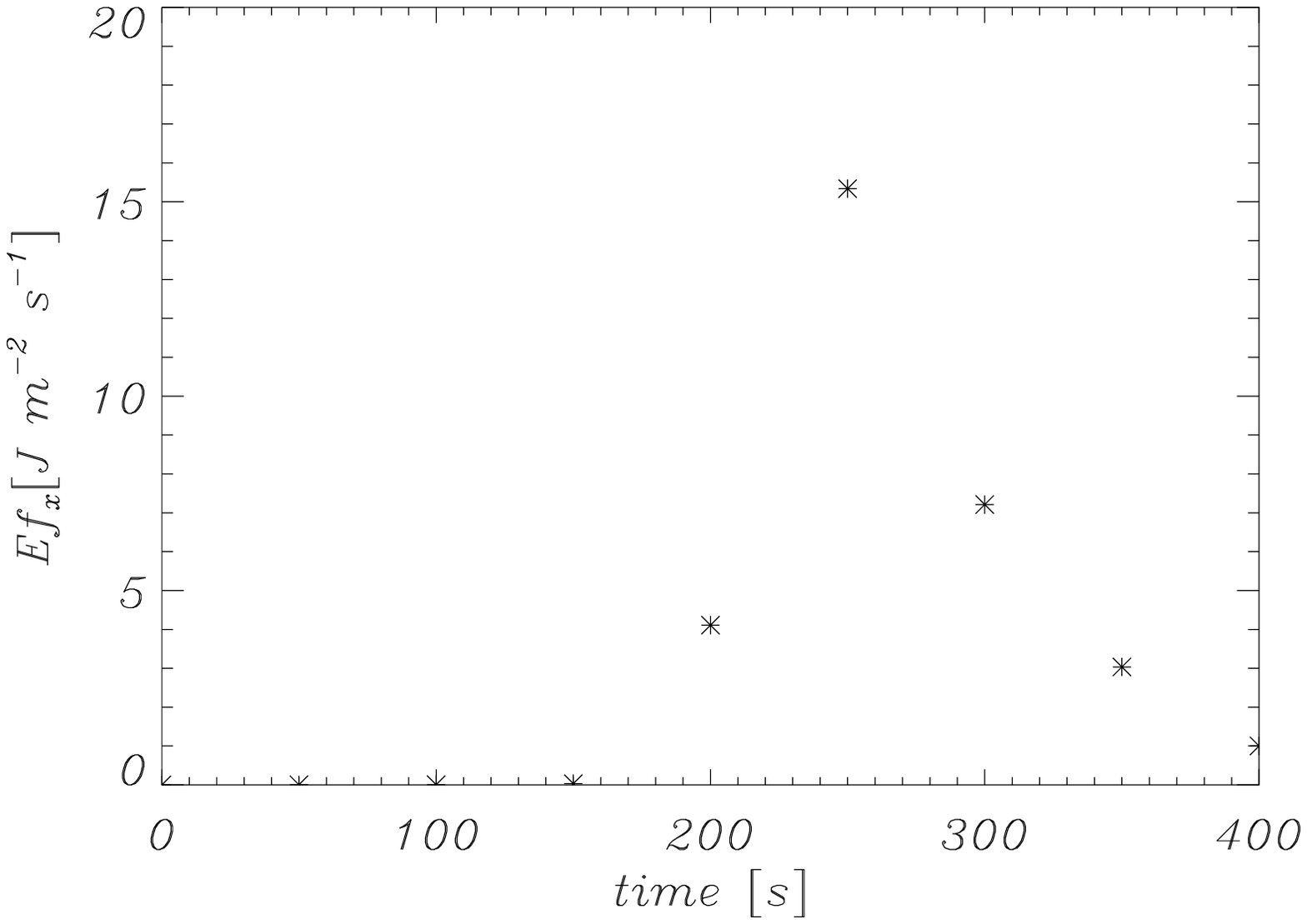}
\includegraphics[scale=0.3, angle=0]{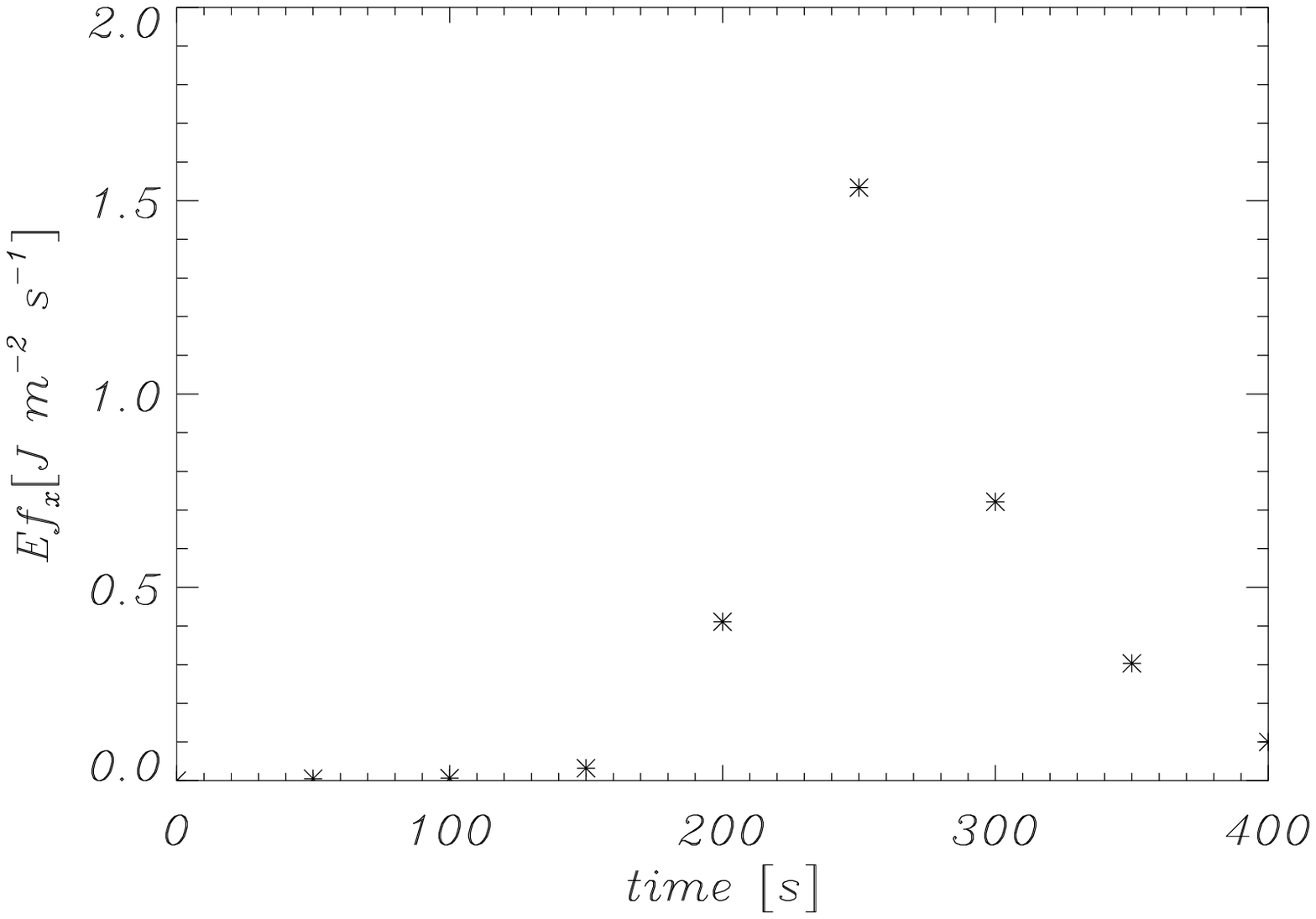}
\includegraphics[scale=0.3, angle=0]{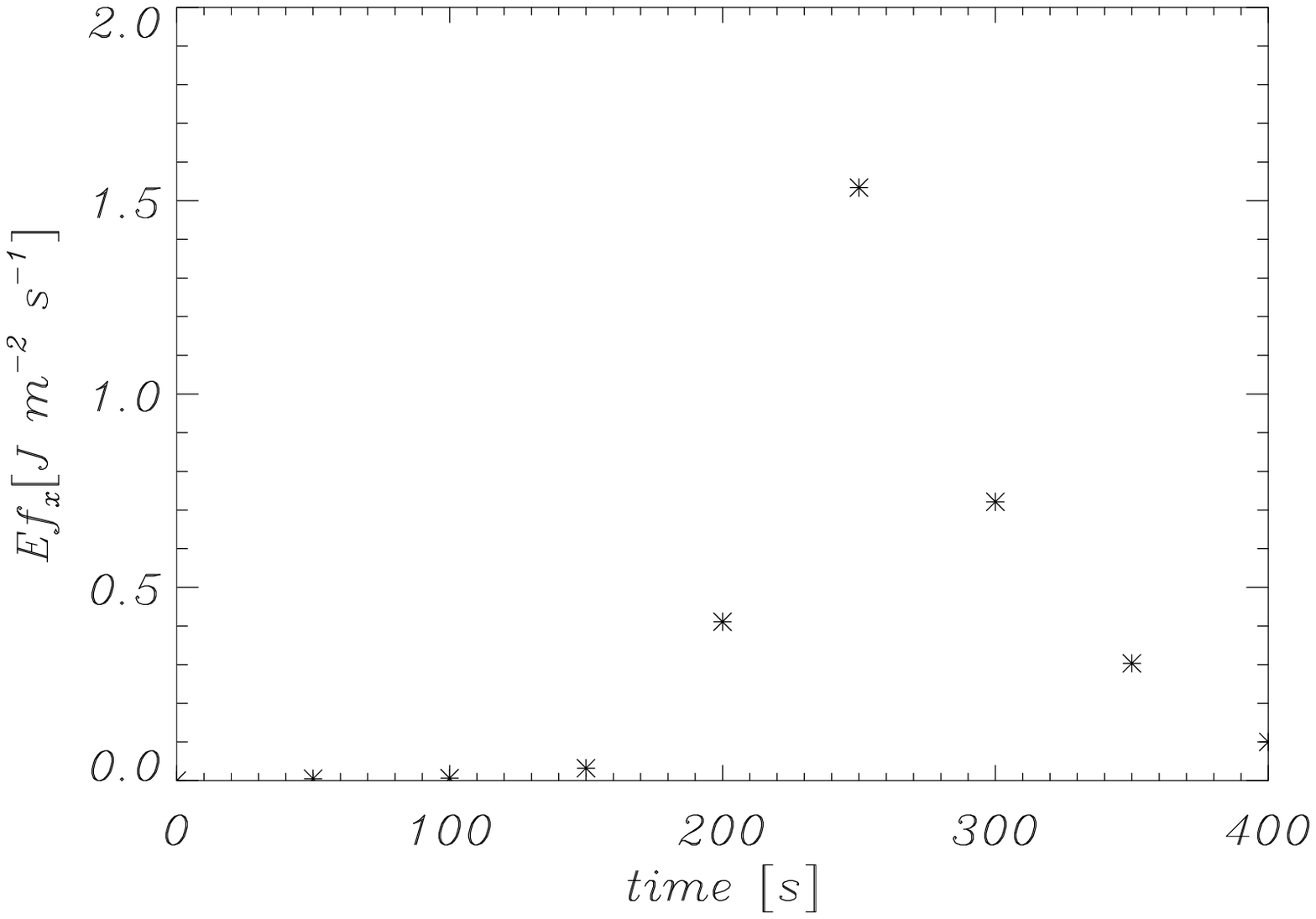}
}
\mbox{
\includegraphics[scale=0.3, angle=0]{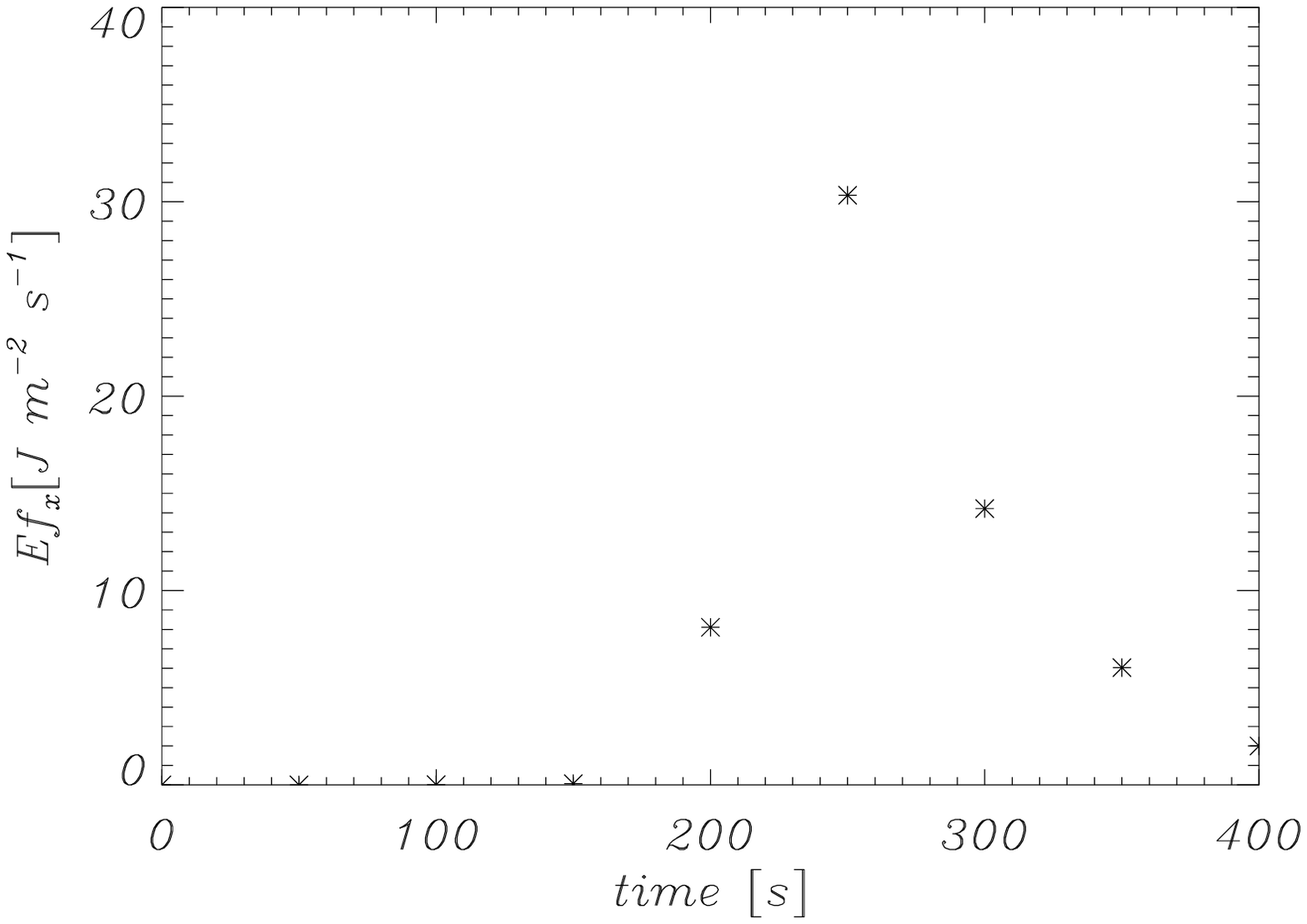}
\includegraphics[scale=0.3, angle=0]{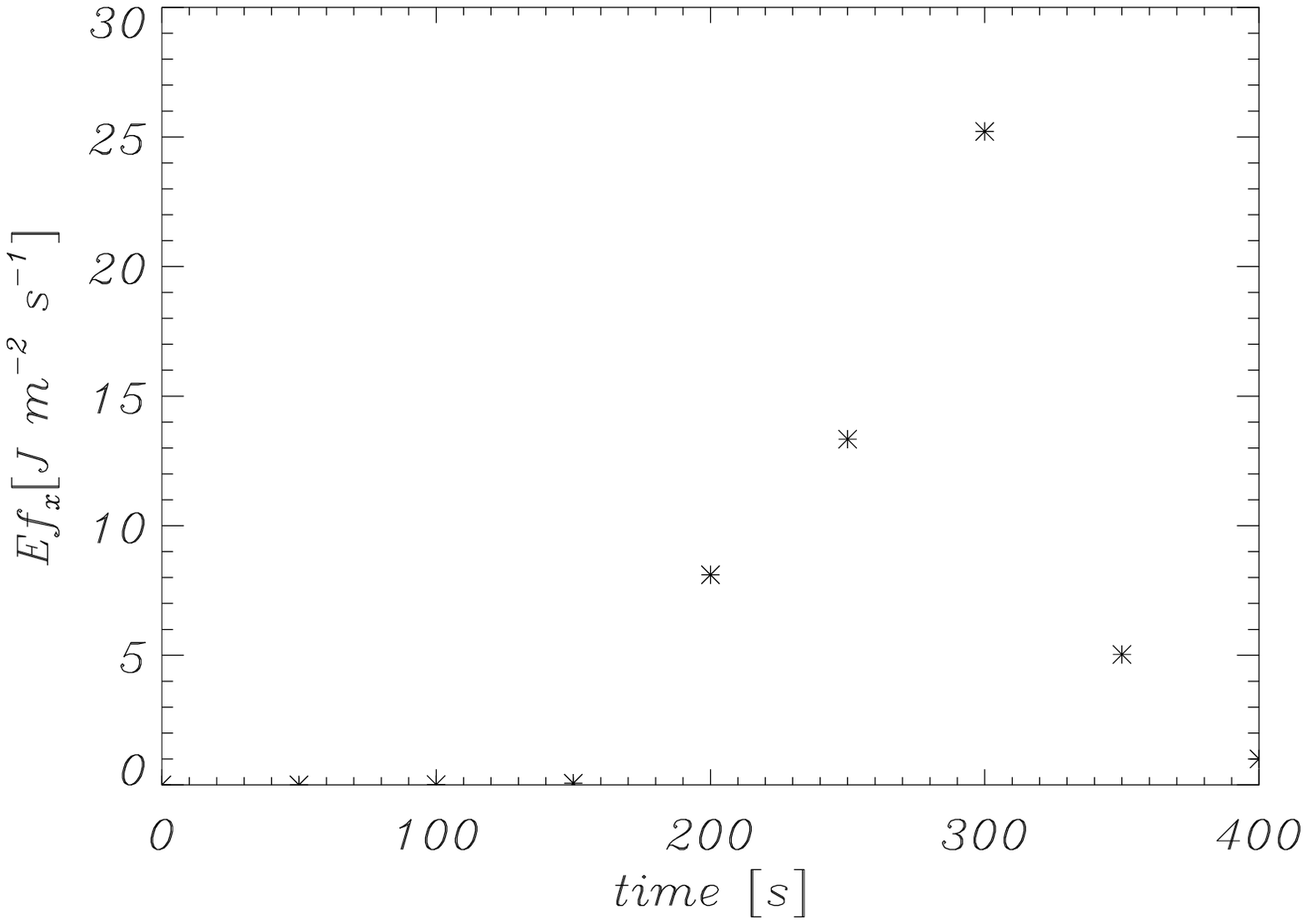}
\includegraphics[scale=0.3, angle=0]{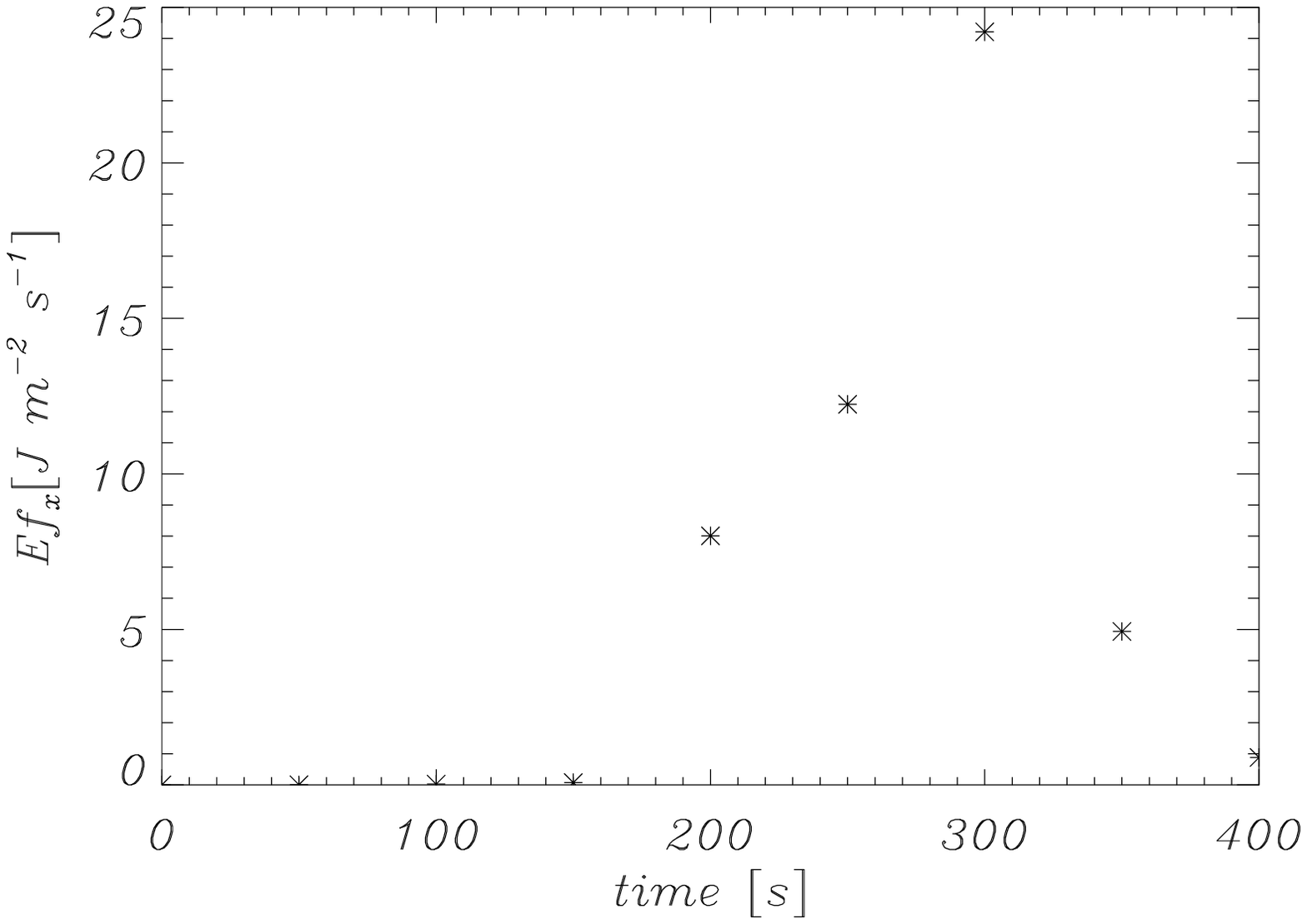}
}
\mbox{
\includegraphics[scale=0.3, angle=0]{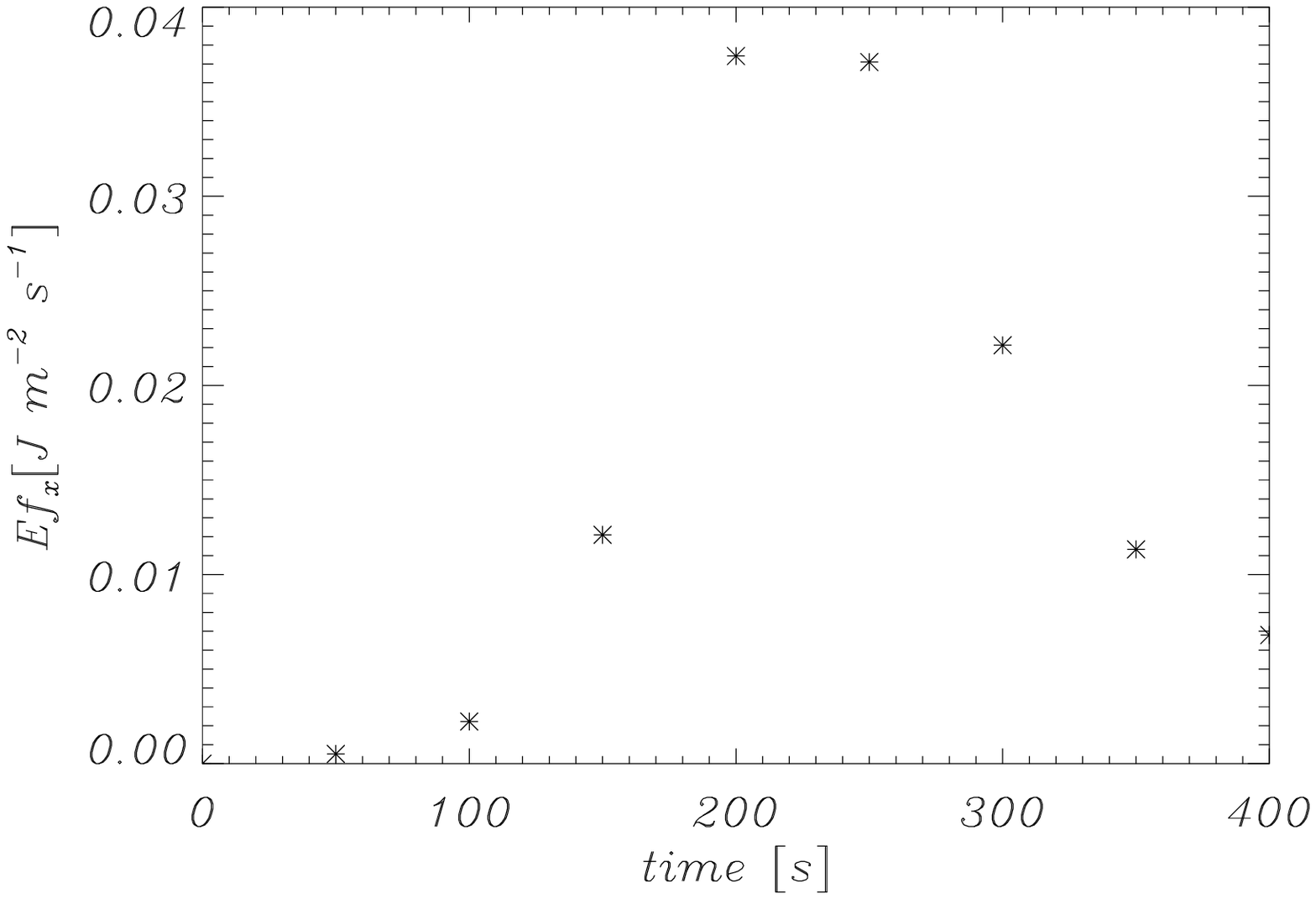}
\includegraphics[scale=0.3, angle=0]{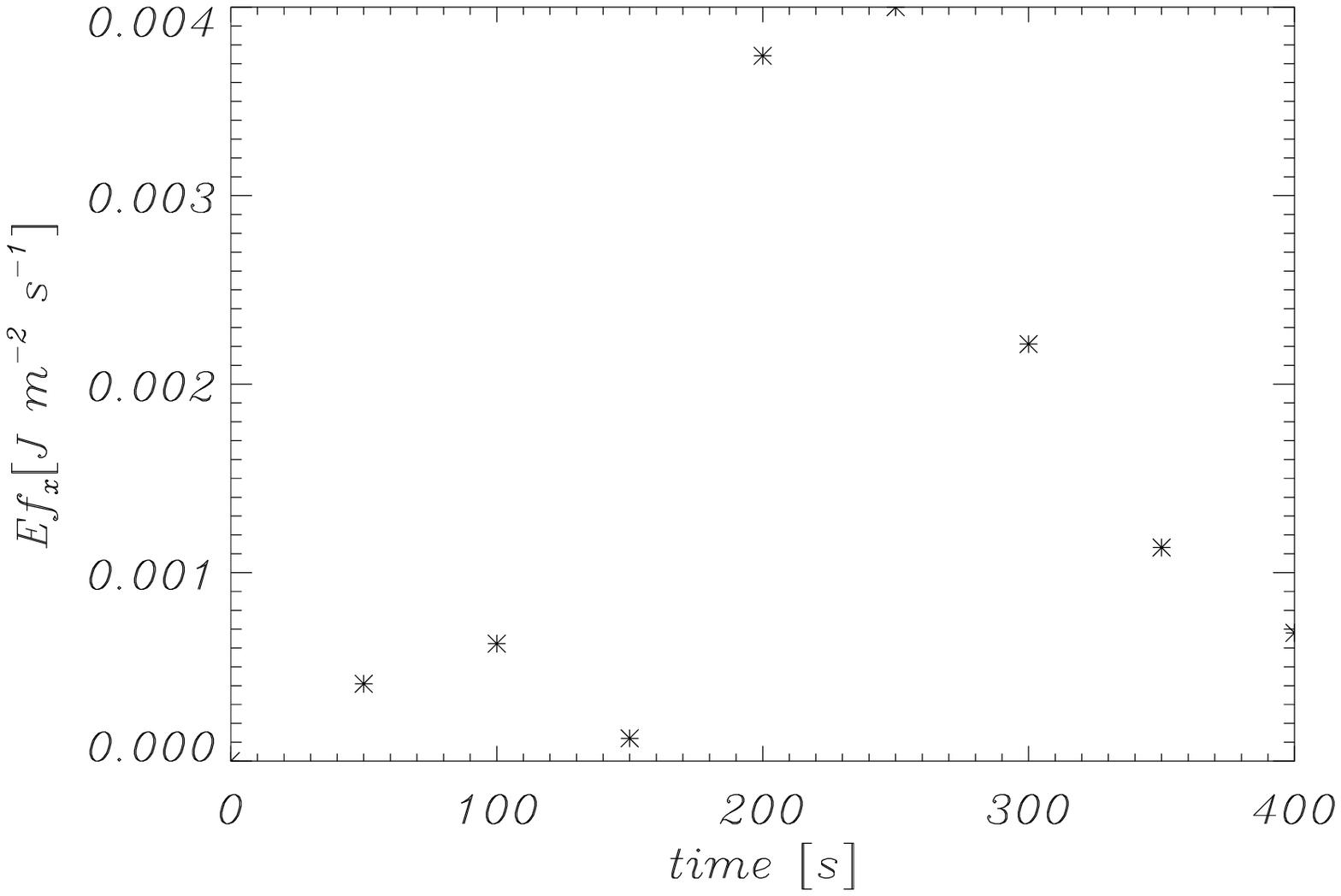}
\includegraphics[scale=0.3, angle=0]{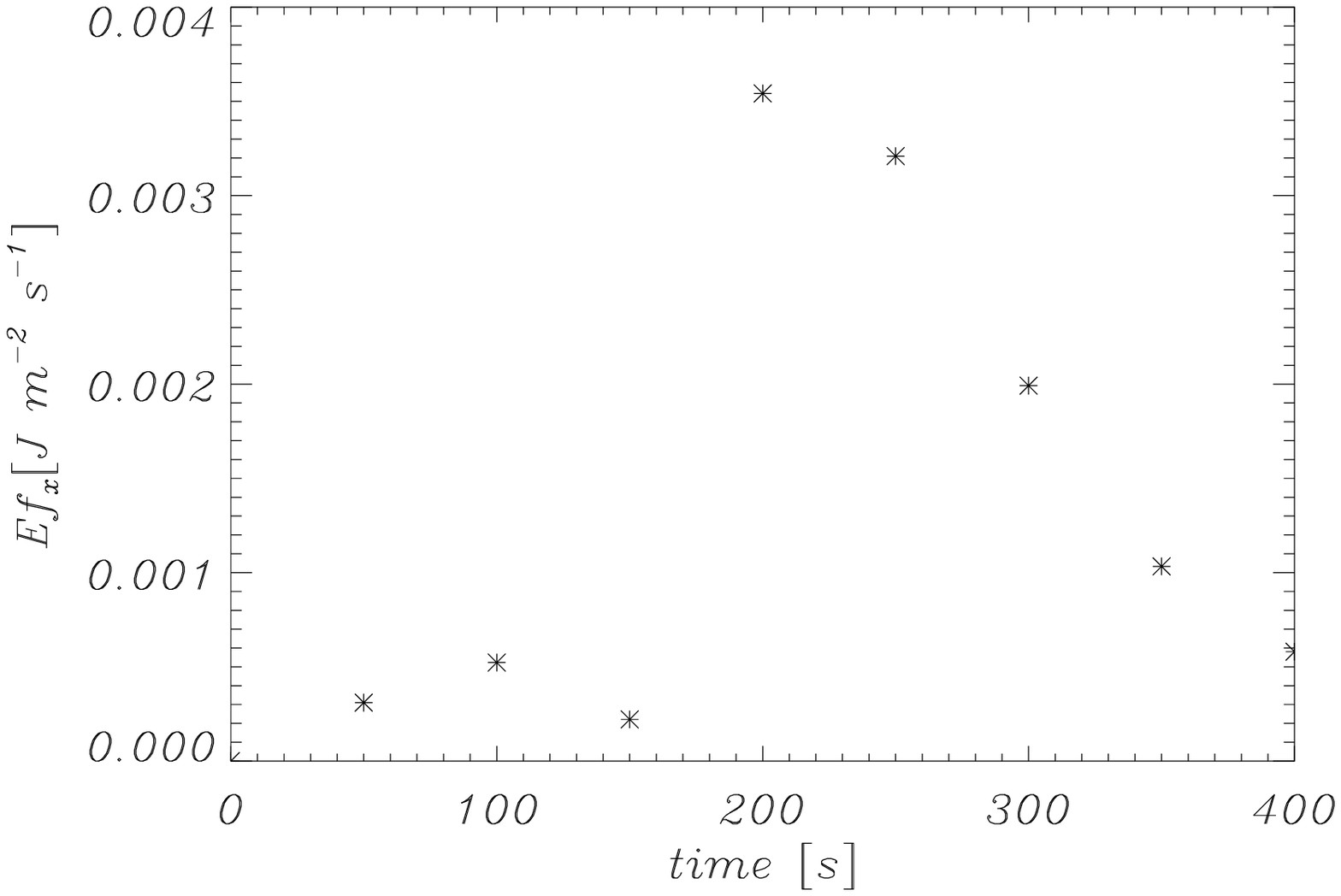}
}
\caption{\small Temporal snapshots of the $x-$, $y-$, and $z-$components 
of the horizontally averaged energy fluxes, $Ef_{\rm x}$, $Ef_{\rm y}$, $Ef_{\rm z}$, evaluated at 
(i) $y=2.5$ Mm (just below the transition region, left columns), 
(ii) $y=3.0$ Mm (slightly above the transition region, middle columns), and 
(iii) $y=3.5$ Mm (in the inner corona, right columns) 
for the case of the horizontal pulse. 
}
\label{fig:hor-Ef}
\end{center}
\end{figure*}

We calculate the horizontally averaged energy fluxes of magnetoacoustic-gravity waves 
generated by the vertical (Fig.~\ref{fig:vert-Ef}) and horizontal (Fig.~\ref{fig:hor-Ef}) pulses. 
These fluxes are evaluated at the levels: 
(i)  $y=2.5$ Mm (just below the transition region, left panels); 
(ii) $y=3.0$ Mm (slightly above the transition region, middle panels); 
and 
(iii) $y=3.5$ Mm (in the inner corona, right panels). 
The components of the energy fluxes are calculated by means of the relations (Vigeesh et al. 2012)
\beqa\label{eq:E_flux}
Ef_{\rm x,z} &\approx& \varrho V_{\rm x,z}^2 \sqrt{c_{\rm s}^2+c_{\rm A}^2}\, , \\
\label{eq:E_fluxy}
Ef_{\rm y}   &\approx& \varrho V_{\rm y}^2   c_{\rm s}\, .
\eeqa
In the case of the vertical pulse, 
it is clear from Fig.~\ref{fig:vert-Ef} that the vertical component of the energy flux is 
about 10$^{3}$ times higher than the horizontal energy flux (compare top and bottom rows). 
Moreover, the vertical energy flux is almost doubled just above the transition region
compared to the same below of it (left-bottom and middle-bottom panels). 
As a result, we conclude that energy is transferred essentially along the vertical direction 
and energy transfer becomes more efficient above the transition region. The vertical component of the energy flux obtained in this way implies that 
the waves could play a significant role in quiet coronal heating requirements
(100-200 J m$^{-2}$ s$^{-1}$, Withbroe and Noyes 1977).

Figure~\ref{fig:hor-Ef} illustrates the horizontally averaged energy fluxes, given by Eqs.~(\ref{eq:E_flux})-(\ref{eq:E_fluxy}) for the case of 
the horizontal pulse. We investigate the horizontal components of the flux ($Ef_{\rm x}$ and $Ef_{\rm z}$) which drop by about one order of magnitude 
at the transition region (compare left and middle columns). As the initial pulse is launched in 
the horizontal velocity, $Ef_{\rm x}$ is about $300$ times larger than $Ef_{\rm z}$ (compare top and bottom raws). 
Note also that $Ef_{\rm y}$ is by a factor of $2$ larger than $Ef_{\rm x}$ just below the transition region 
(left-top and left-middle panels). Although the initial driver was horizontal, at higher altitudes $Ef_{\rm y}$ drops twice and it is about $10$ times larger 
than $Ef_{\rm x}$ in the inner corona (compare middle and top rows). As $Ef_{\rm y}$ is much larger than $Ef_{\rm x}$ we infer 
that energy transport occurs essentially along the vertical direction. Comparing the values of the energy fluxes obtained for 
the two types of the driver we can conclude that the amount of energy that can be transported into the solar corona as a result of 
a perturbation driven vertically is an order of magnitude larger than the same obtained by a horizontal driver of the same strength.

\begin{figure}
\begin{center}
\includegraphics[width=6.00cm, angle=0]{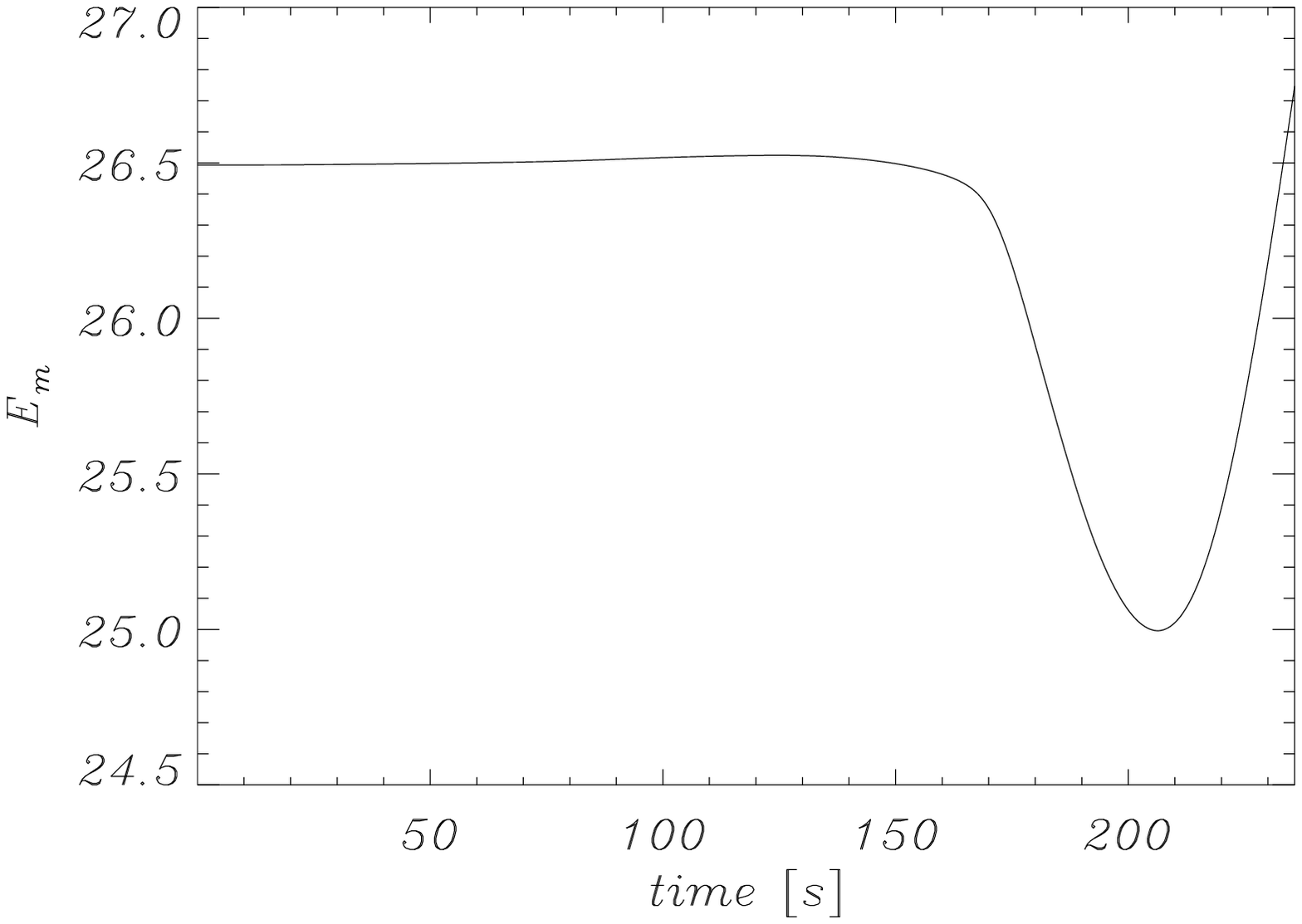}
\includegraphics[width=6.00cm, angle=0]{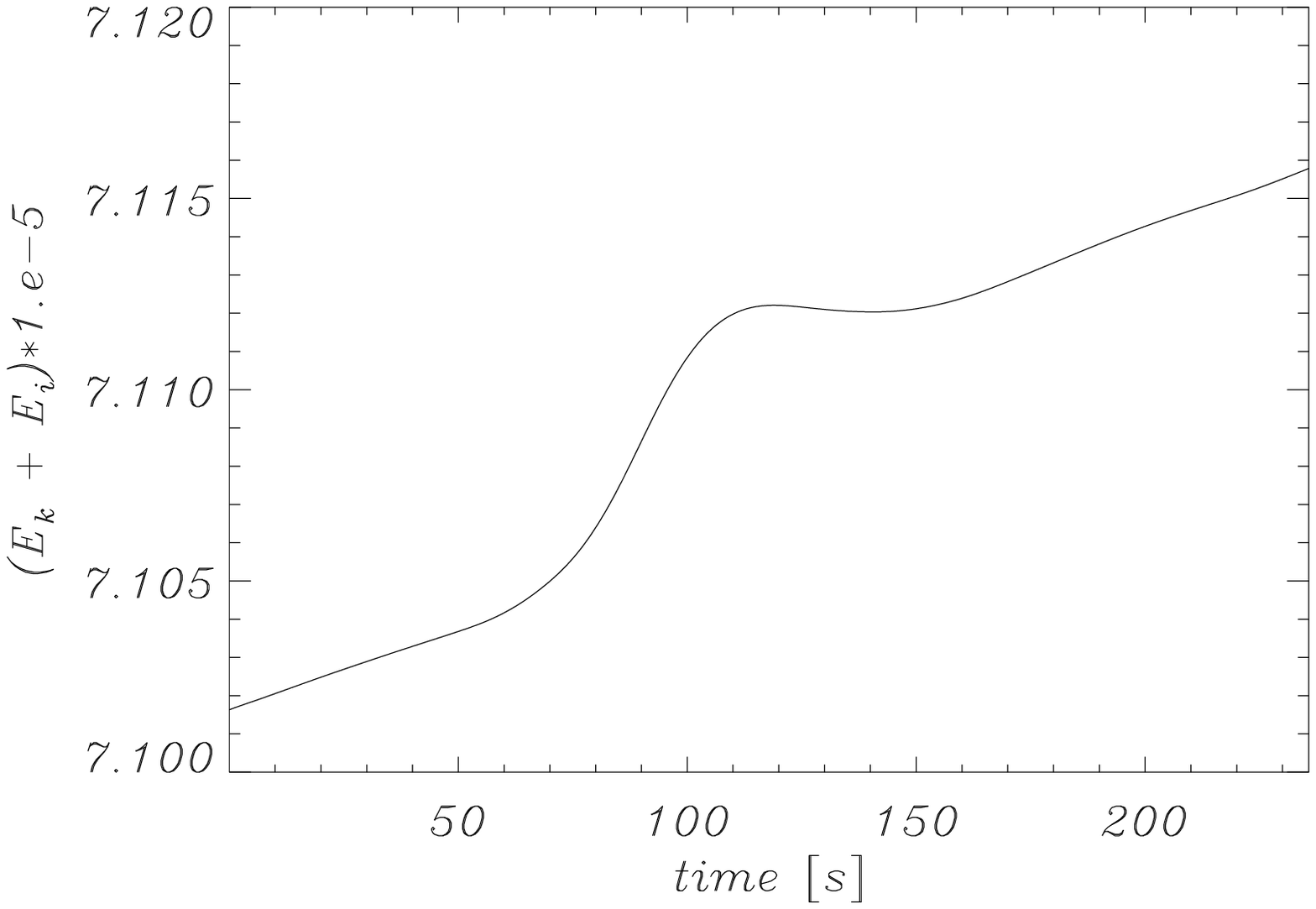}
\caption{\small 
Time-history of magnetic (top) and sum of kinetic and internal (bottom) energies in the simulation box for 
for the case of the vertical perturbation with $s_{\rm x}=0$, $s_{\rm y}=1$ in Eq.~(\ref{eq:perturb}). 
}
\label{fig:vert-B:mech-ener}
\end{center}
\end{figure}

Figure~\ref{fig:vert-B:mech-ener} shows the time-signatures of magnetic (top) and sum of the kinetic and internal (bottom) energies in the simulation box for 
the case of vertical perturbation with $s_{\rm x}=0$, $s_{\rm y}=1$ in Eq.~(\ref{eq:perturb}). 
As the initial magnetic field is vertical, uniform and therefore potential, it has the lowest possible energy state. 
In reconnection, we would expect an accumulation of energy in magnetic fields (similar to what is happening between $50$ and $150$ seconds) 
and then its decrement to the initial value of the normalized magnetic energy which is $26.5$ according to the plot when the magnetic fields reconnect. 
Simultaneously, there should be a local rise in the sum of kinetic and internal energies (bottom). 
The drop in the energy down to $25.0$ (less than the initial energy state) is most probably due to 
the magnetic field leaving the simulation box through the side boundaries which are transparent. The following rise of the magnetic 
energy is probably advection of the magnetic fields into the box through the same boundaries. In general, it looks more like a start of an oscillation within the 
simulation box. 
A constant rise of the sum of kinetic and internal energies (bottom) 
smoothly rises by a significant value far above normal numerical noise levels. 

\begin{figure}
\begin{center}
\includegraphics[width=7.00cm,height=8.50cm,angle=0]{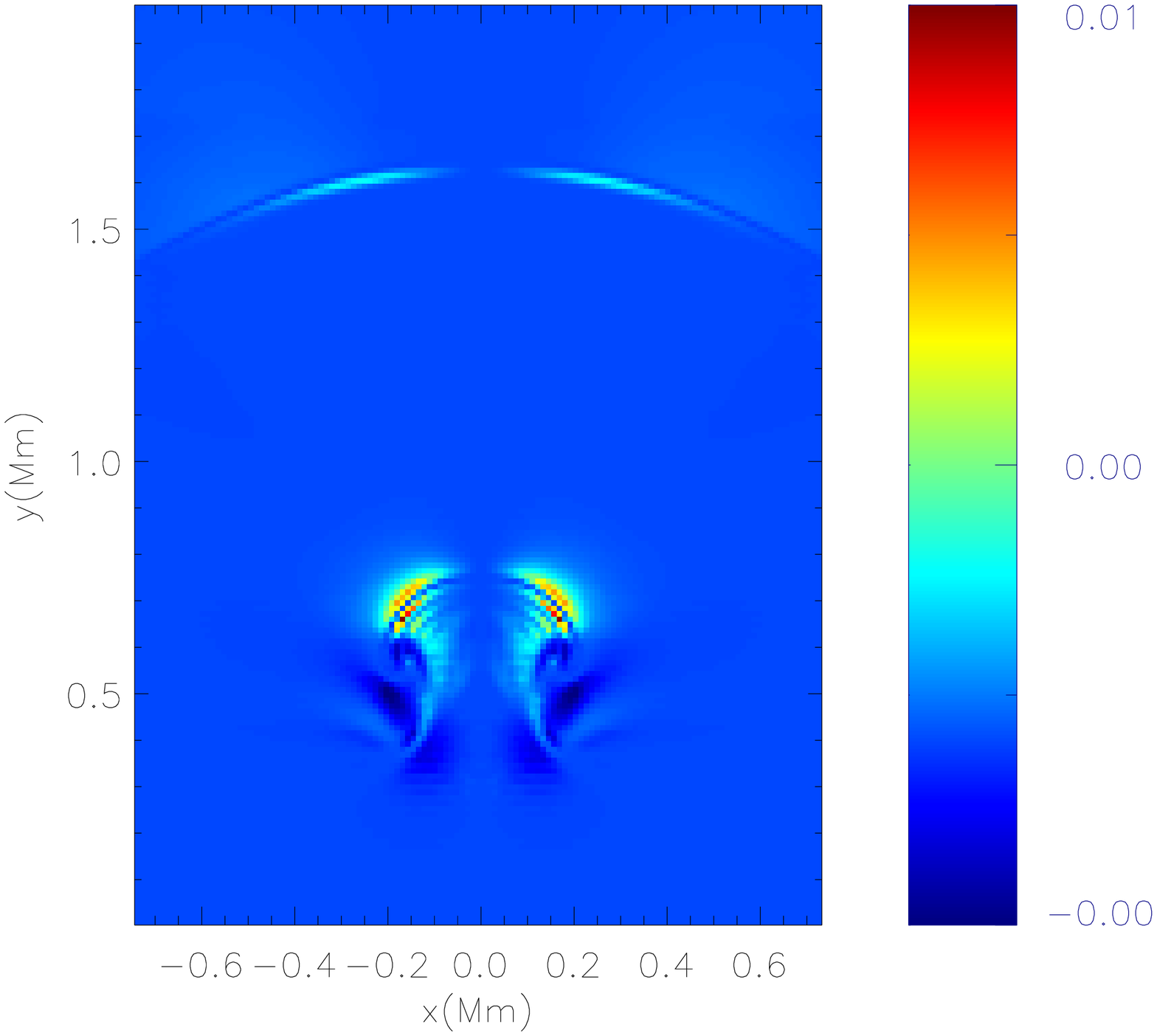}
\includegraphics[scale=0.45,angle=0]{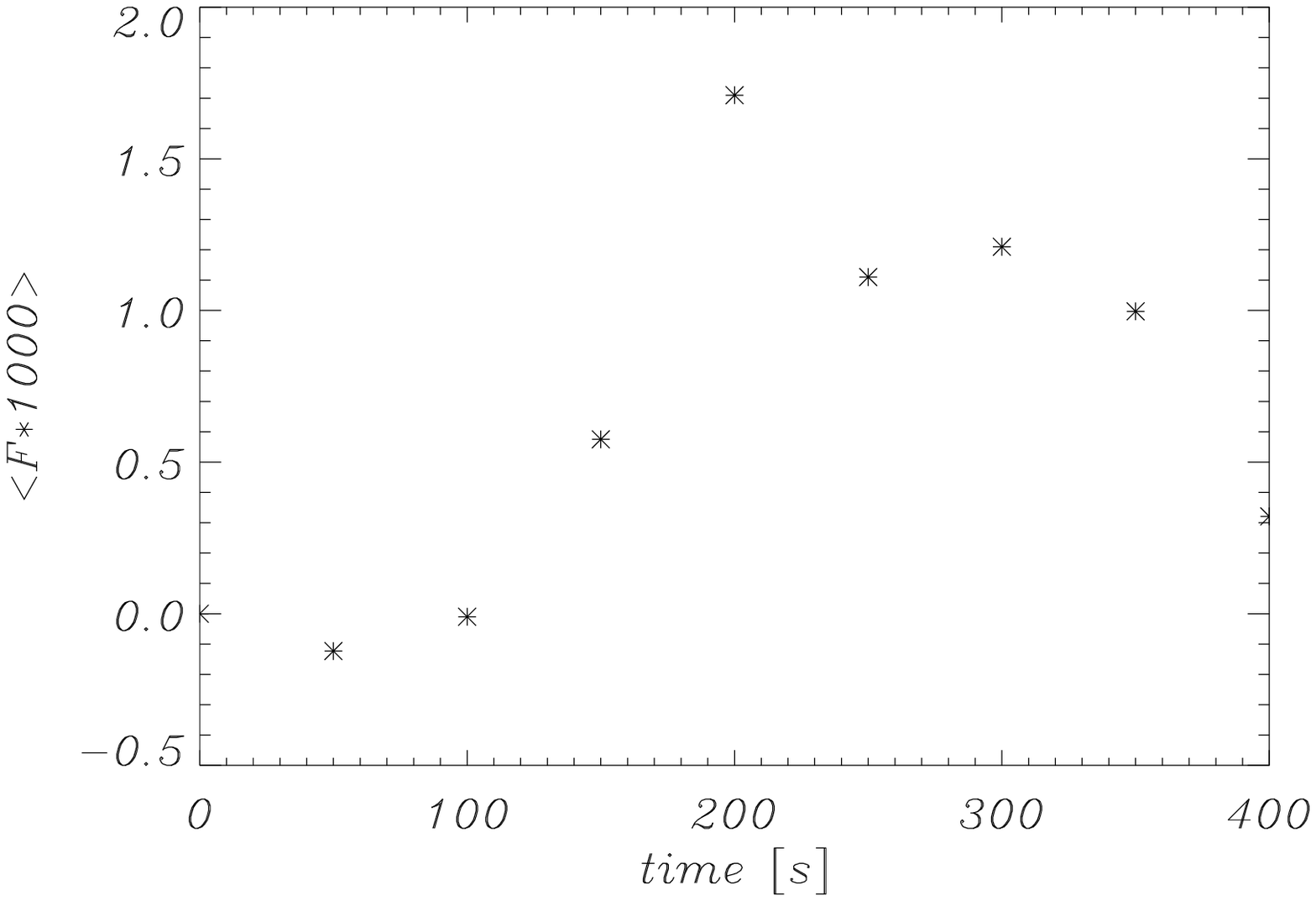}
\caption{\small 
Vertical component of the Poynting flux at $t=200$ s 
(top) and 
horizontally averaged Poynting flux 
at $y=0.75$ Mm vs. time 
(bottom)
for the case of the vertical perturbation with $s_{\rm x}=0$, $s_{\rm y}=1$ in Eq.~(\ref{eq:perturb}). 
}
\label{fig:vert-B:Poynt}
\end{center}
\end{figure}

Figure~\ref{fig:vert-B:Poynt} illustrates the vertical component of the Poynting flux (top) which is defined as 
\beq
{\bf F} = \frac{1}{\mu} {\bf B}\times ({\bf V}\times {\bf B})\, .
\eeq
Horizontally averaged Poynting flux, $<F>$, at the level $y=0.75$ Mm is displayed in bottom panel. 
It is discernible that $<F>$ is essentially positive in agreement with the numerical findings of Shelyag et al. (2012). 
%
%
\section{Summary}
%
This paper presents the stringent 3D numerical modeling of 
the solar atmosphere with an adaptation of the FLASH code, which demonstrates the feasibility of fluid simulations in obtaining 
quantitative features in the weakly magnetized and gravitationally stratified solar atmosphere. 
These simulations reveal the generation and propagation of MHD-gravity waves and vortices in the solar atmosphere, 
which are impulsively excited by an initial localized pulse in a velocity component 
that is either horizontal or vertical. 
In both cases the initial pulse triggers vortices with their 
axis oriented either along the $y$- or $x$-direction depending on 
the vertical or horizontal perturbations. 
These vortices develop in time into more complex structures which are typical for turbulence. 
It is apparent that large-sized eddies cascade into small-scale vortices. 
The latter are prone to dissipation which may transfer mechanical energy 
into heat and therefore they may contribute to a deposition of thermal energy in higher layers of the solar atmosphere. 
However, the detailed study of the dissipation of energy carried by such vortices and associated waves is 
out of the scope of the present paper and will be undertaken in our future works. 

The convectively driven initial perturbations in form of the photospheric vortices or 
similar kind of plasma rotatory motions can launch their responses in form of the 
plasma dynamics into the upper atmosphere at diverse spatio-temporal scales to channel
the mass and energy and thus to heat the solar corona locally. 

Our presented 3D numerical simulations contribute to understanding such physical conditions 
in the model solar atmosphere with realistic temperature and related conditions within the strongly magnetized 
flux-tubes rooted at the photosphere and fanning out up to inner corona. Our model mimic the formation of the 
vortices, magnetoacoustic--gravity waves, and their responses into the chromosphere, transition region, and inner corona.
The evolved model of vortices in the 3D atmosphere with their upper atmospheric responses can be considered 
very similar to the dynamics as swirling motions that are considered to be the potential candidates to channel the energy in localized atmospheric heating. 
Our model, therefore, can be considered as a potential reference to the various observed and 
recurring swirling motions into the solar atmosphere. 
Temporal evolution of energy fluxes over the three directions revealed that irrespective of the type of the driver, 
by the time the perturbation reach the transition region, the dominant direction of energy flow will be always in the vertical direction.

In summary, we initiated three-dimensional numerical models of 
energy transport from the photosphere into the corona in form of vortices and mag\-ne\-to\-acou\-stic-gravity waves. 
This 3D numerical model of solar vortices, associated waves, and their responses in the 
upper atmosphere may shed some light on the energy transport processes that can further cause the chromospheric and coronal heating in the case 
of their dissipation. The detailed physical analyses of MHD-gravity waves and the various type of vortices and
their coronal/transition region responses as well as their vis-to-vis comparison 
with the observations, will be the part of our forthcoming work. 

{\bf Acknowledgments.} 
We thank the reviewer for his/her valuable suggestions
that improved our manuscript considerably. 
The author (KM) expresses his thanks to Kamil Murawski for his assistance in drawing the
numerical data. 
AKS acknowledges the support of DST-RFBR (INT/RFBR/P-117) project, as well as
patient encouragement from Shobhna Srivastava in his solar research. IB acknowledges the
financial support by NSF Hungary (OTKA, K83133).
The software used in this work was in part developed by the DOE-supported ASCI/Alliance Center 
for Astrophysical Thermonuclear Flashes at the University of Chicago. 
%
The 2D and 3D visualizations of
the simulation variables have been carried out using respectively the IDL (Interactive Data Language) and 
VAPOR (Visualization and Analysis Platform) software packages.
The computational resources were provided by 
the ``HPC Infrastructure for Grand Challenges of Science and Engineering" Project, 
co-financed by the European Regional Development Fund under 
the Innovative Economy Operational Program. 



\end{document}